\documentclass[a4paper]{jpconf}
\usepackage{graphicx,eufrak}
\usepackage[numbers,sort&compress]{natbib}
\usepackage{wasysym}\usepackage[usenames]{xcolor}
\usepackage{bm}
\begin{document}
\title{The Yang--Mills vacuum wave functional thirty-five years later${}^\ast$}

\author{\v{S}tefan Olejn{\'\i}k}

\address{Institute of Physics, Slovak Academy of Sciences, SK--845 11 Bratislava, Slovakia}

\ead{stefan.olejnik@savba.sk}

\begin{abstract}
%Beside anniversaries of the important discoveries celebrated at the \emph{DISCRETE~2014\/} Symposium, there is another worth to mention: 
The first paper attempting direct calculation of the Yang--Mills vacuum wave functional was published by Greensite  %~\cite{Greensite:1979yn}. 
 in 1979. I review some recent results of the determination of the vacuum wave functional in Monte Carlo simulations of SU(2) lattice gauge theory.
\end{abstract}

\begin{figure*}[b!]
\footnotesize${}^\ast$~~Based on talks presented at the \textit{Fourth Symposium on Prospects in the Physics of Discrete Symmetries, DISCRETE 2014,\/} King's College, London, UK, December 2-6, 2014, and (under the title \textit{Measurement of the Yang--Mills vacuum wave functional in lattice simulations\/}) at the \textit{4th Winter Workshop on Non-Perturbative Quantum Field Theory}, INLN, Sophia Antipolis, France, February 2-5, 2015. Abridged versions will be published in proceedings of these conferences.
\end{figure*}
\section{Introduction}

The \emph{DISCRETE~2014\/} Symposium commemorated a number of important events that had shaped modern physics. Maxwell presented his theory of electromagnetism to the Royal Society 150~years ago; non-abelian gauge theories were proposed by Yang and Mills 60 years ago; the Brout--Englert--Higgs mechanism, the quark model, and Bell inequalities are 50 years old; CP violation was experimentally discovered 50 years ago as well. In my talk I would like to draw attention to another anniversary: In February 1979, the first paper trying to calculate the ground-state wave functional of Yang--Mills (YM) theory was submitted to \emph{Nuclear Physics\/} by Jeff Greensite~\cite{Greensite:1979yn}. 
Thirty-five years have passed since then, but the problem still defies satisfactory solution.    

\paragraph{\textbf{Formulation of the problem}}
The vacuum wave functional (VWF) $\mathbf{\Psi}_0$ of quantum chromodynamics in the Schr\"odinger representation depends on quark fields of six flavours with three colours, each represented by a Dirac four-component bispinor, and on eight four-vector gluon fields -- this is altogether 104 fields at each point in space (not taking constraints from gauge invariance into account). This is a formidable object from both mathematical and practical point of view. To simplify the problem, one can reduce the number of colours from three to two, omit quarks, discretize space (\textit{i.e.\/} formulate the theory on a lattice), and eventually go to lower-dimensional spacetime. One can hope \cite{Feynman:1981ss} that the resulting model captures at least gross features of the full theory, in particular information on the mechanism of colour confinement.

Omitting quarks, the SU(2) YM Schr\"odinger equation in $(d+1)$ dimensions  in temporal gauge looks very simple:
\begin{equation}\label{YMSchR}
\hat{\cal H}\mathbf{\Psi}[A]=
\int d^d x\left[-\frac{1}{2}
\frac{\delta^2}{\delta A_k^a(x)^2}+\frac{1}{4}F_{ij}^a(x)^2\right]\mathbf{\Psi}[A]=
E\mathbf{\Psi}[A].
\end{equation}
Physical states are simultaneously required to satisfy Gau\ss' law:
\begin{equation}\label{GaussLaw}
\left(\delta^{ac}\partial_k+
g\epsilon^{abc}A_k^b\right)
\frac{\delta}{\delta A_k^c}\mathbf{\Psi}[A]=0.
\end{equation}

\paragraph{\textbf{A few well-known facts}}
If we set the gauge coupling $g$ to 0, the Schr\"odinger equation (\ref{YMSchR}) reduces to that of (three copies of) electrodynamics and its ground-state solution is known to be \cite{Wheeler:1962aa}:
\begin{equation}\label{PsiQED}
\mathbf{\Psi}_0[A]\;
{\stackrel{g=0}{=}}\;
{\cal{N}}\exp\left[-{\textstyle\frac{1}{4}}{\displaystyle\int} d^dx\;d^dy\; 
F^a_{ij}(x)\;
\left(\frac{\delta^{ab}}{\sqrt{-\mathbf{\Delta}}}\right)_{xy}F^b_{ij}(y)\right].
\end{equation}
Due to (\ref{GaussLaw}), the ground state must be gauge-invariant. The simplest form consistent with (\ref{PsiQED}) is
\begin{equation}\label{PsiGI}
\mathbf{\Psi}_0[A]\;=\;
{\cal{N}}\exp\left[-{\textstyle\frac{1}{4}}\int d^dx\;d^dy\; 
F^a_{ij}(x)\;
{\mathfrak{K}}^{ab}(x,y)\;F^b_{ij}(y)\right],
\end{equation}
where $\mathfrak{K}$ is some adjoint-representation kernel that reduces, in the limit $g\to0$, to $(-\mathbf{\Delta})^{-1/2}$.

\begin{figure}[t!]
\fbox{\includegraphics[width=17.5pc]{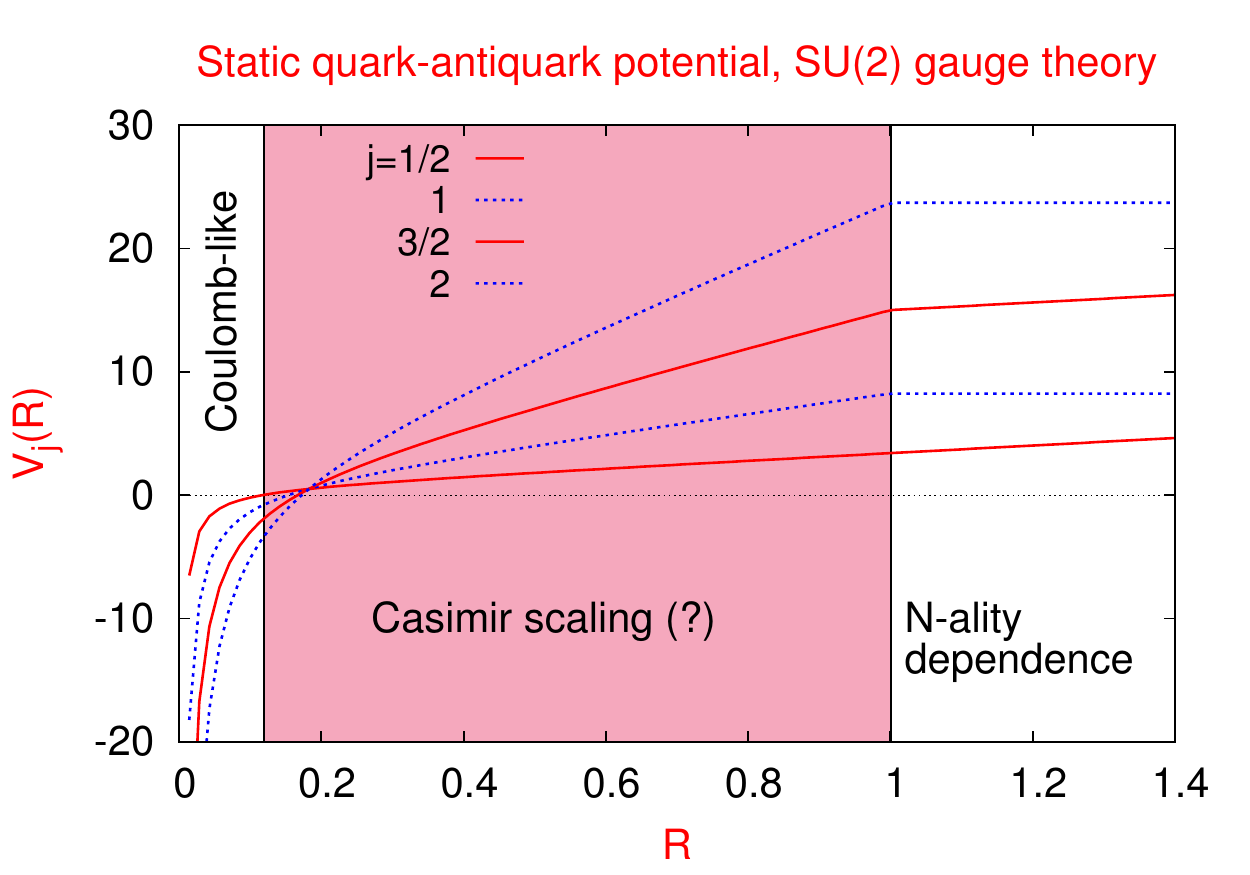}}\hspace{1.5pc}%
\begin{minipage}[b]{18pc}\caption{\label{potentials}Sketch of static potentials in SU(2): At short distances the potentials are Coulomb-like. At large distances, colour charges of higher-representation sources can be screened by gluons. Potentials of integer representations become asymptotically flat, while asymptotic string tensions for all representations with half-integer $j$ are the same as for $j=1/2$. Casimir scaling of string tensions is expected only at intermediate distances, at least in some models of confinement.}
\end{minipage}
\end{figure}

 At long-distance scales, one expects the VWF to be the state of \textit{magnetic disorder} \cite{Halpern:1978ik,Greensite:1979yn,Kawamura:1996us}:
 \begin{equation}\label{PsiDR}
 \mathbf{\Psi}_0[A]\approx
{\cal{N}}\exp\left\{-{\textstyle\frac{1}{4}}\;\mu\int d^dx\; 
F^a_{ij}(x)\;F^a_{ij}(x)\right\}.
 \end{equation}
This is also called the \textit{dimensional-reduction} (DR) form. With such a VWF,  the computation of a spacelike loop in $(d+1)$ dimensions reduces to the calculation of a Wilson loop in YM theory in $d$ (euclidean) dimensions. If the vacuum were of the DR form
for YM theories in both $(3+1)$ and $(2+1)$ dimensions, then these would be confining, since the theory in 2 euclidean dimensions exhibits the area law. However, this cannot be the whole truth, for various reasons. Via DR, one gets \textit{e.g.\/} the area law for colour charges from \textit{all\/} representations $r$ of the gauge group, and string tensions $\sigma_r$ of the corresponding potentials are proportional to eigenvalues of the $r$-th~quadratic Casimir operator. Approximate \textit{Casimir scaling\/} is observed at short and intermediate distances, but at large distances, due to \textit{colour screening\/}, string tensions depend on the $N$-ality of the representation (see figure \ref{potentials} for illustration).  

%\begin{center}
%\end{center}

\paragraph{\textbf{Approaches to the problem}}
A number of different strategies were adopted in attempts to determine the YM vacuum wave functional (see \cite{Greensite:2011pj} for a~more extensive set of references):
\begin{enumerate}
\item
Strong-coupling expansion of the VWF \cite{Greensite:1979ha,Guo:1994vq}.
\item
Weak-coupling expansion of the VWF \cite{Hatfield:1984dv,Krug:2013yq,Krug:2013cka,Krug:2014oma}.
\item
Formulation of the theory in cleverly selected variables and expansion of the VWF in terms of these new variables \cite{Karabali:1998yq,Agarwal:2007ns,Leigh:2005dg,Leigh:2006vg,Freidel:2006qy,Freidel:2006qz,Schneider:2011zz}.
\item
Variational Ansatz for the VWF in a certain gauge; determination of its parameters by minimizing the
expectation value of the YM hamiltonian (see \textit{e.g.\/} \cite{Szczepaniak:2001rg,Feuchter:2004mk}).
\item
Guess of an approximate form of the VWF and tests of its consequences (see \cite{Samuel:1996bt}, \cite{Greensite:2007ij,Greensite:2010tm,Greensite:2013rva}).
\end{enumerate}

\section{Flatland: A romance of two dimensions%
\footnote{This section's heading paraphrases the title of a novella \cite{Abbott:1884aa} by English schoolmaster and theologian Edwin Abbott Abbott, writing pseudonymously as \textit{A Square}, which describes a two-dimensional world inhabited by geometric figures. The book was published in 1884 by a publishing house at that time located in Essex Street, less than half a kilometre from King's College.}}\label{flatland}

I will start with the problem in the simplest setting: with SU(2) Yang--Mills theory in \textit{``Flatland''\/}, \textit{i.e.\/} in three spacetime dimensions. That model deserves attention not only because it is more tractable than the realistic case, but also due to its relevance to the high-temperature phase of chromodynamics in four dimensions.\footnote{Another good justification -- not scientific, but utterly human -- for investigating the model in lower dimensions was given by Feynman in his lecture at the 1981 EPS--HEP Conference in Lisbon \cite{Feynman:1981zz}: \textit{``Of course, understanding something in $2+1$ dimensions does not imply that you understand it in $3+1$ dimensions, but, after all I have lots of time, I have tenure, so I can do whatever I want.''} I would avoid writing such a justification into a grant proposal, but find it nevertheless very reasonable.}

\paragraph{\textbf{Hints on the form of the VWF}}
In the lattice formulation, a systematic strong coupling expansion of the YM VWF is of the form $\mathbf{\Psi}_0={\cal{N}}\exp(-R[U])$, where the function $R$ in the exponent is an expansion in terms of closed loops, products of link matrices $U$ along closed contours on the lattice \cite{Greensite:1979ha}. Guo, Chen and Li \cite{Guo:1994vq} computed the first few terms of this expansion and showed that for slowly varying fields they organize themselves into the following series\footnote{The trace in this symbolic expression includes sums over colour indices and lattice sites.}:
\begin{equation}\label{Guo}
R[U]\propto\mu_0\mbox{Tr}\;[\mathbf{B}^2]-\mu_1\mbox{Tr}\;[\mathbf{B}(-{\cal D}^2)\mathbf{B}]+\dots
\end{equation}
Here $\mu_0$ and $\mu_1$ are functions of the lattice spacing $a$ and the coupling constant $g$, $\mathbf{B}=\mathbf{F}_{12}$ is the colour magnetic field strength, and ${\cal{D}}^2={\cal{D}}_k\cdot{\cal{D}}_k$ is the adjoint covariant laplacian, where ${\cal{D}}_k[\mathbf{A}]$ denotes the covariant derivative in the adjoint representation. The first term of (\ref{Guo}) corresponds to the dimensional-reduction vacuum wave functional (\ref{PsiDR}). One can imagine that an expansion of the form (\ref{Guo}) might come from the wave functional  
\begin{equation}\label{PsiGIwithD2}
\mathbf{\Psi}_0[A]\;=\;
{\cal{N}}\exp\left\{-{\textstyle\frac{1}{2}}\int d^2x\;d^2y\; 
B^a(x)\;
{\mathfrak{K}}^{ab}_{xy}\left[-{\cal{D}}^2\right]\;B^b(y)\right\},
\end{equation}
with the kernel $\mathfrak{K}$, introduced in (\ref{PsiGI}), being a functional of the adjoint covariant laplacian.

A similar hint may be deduced also from the approach of Karabali \textit{et al.\/}~\cite{Karabali:1998yq}. They combine two components of
the gauge potential into complex-valued fields $\{\mathbf{A},\bar{\mathbf{A}}\}=\frac{1}{2}\left(\mathbf{A}_1\pm i\mathbf{A}_2\right)$
and introduce new variables, a matrix-valued field $\mathbf{M}\in\mbox{SL}(N,{\cal{C}})$, related to $\mathbf{A},\bar{\mathbf{A}}$ via
\begin{equation}\label{KKNvariables}
\mathbf{A}=-\left(\partial_z\mathbf{M}\right)\mathbf{M}^{-1},\qquad
\bar\mathbf{A}=\mathbf{M}^{\dagger-1}\left(\partial_{\bar{z}}\mathbf{M}^\dagger\right),\qquad\mbox{where }
\{z,\bar{z}\}=x_1\pm i x_2.
\end{equation}
$\mathbf{M}$ transforms covariantly, $\mathbf{M}\to\mathbf\Omega\mathbf{M}$, under a gauge transformation $\mathbf\Omega$, and is used to define gauge-invariant variables:
\begin{equation}\label{KKNvariables}
\mathbf{H}=\mathbf{M}^\dagger\mathbf{M},\qquad
{J}^a=\mbox{Tr}\left(T^a(\partial_z\mathbf{H})\mathbf{H}^{-1}\right),
\end{equation}
through which one expresses the hamiltonian, inner products of physical states, and the VWF.

Karabali \textit{et al.\/} further show that the part of the VWF bilinear in variables $J^a$, when expressed through original colour magnetic fields, takes on the form: 
\begin{equation}\label{PsiKKN}
\mathbf{\Psi}_0[A]\approx
{\cal{N}}\exp\left[-{\textstyle\frac{1}{2}}\int d^2x\;d^2y\; 
B^a(x)\;
\left(\frac{1}{\sqrt{-\mathbf{\Delta}+m^2}+m^2}\right)_{xy}
\;B^a(y)\right].
\end{equation}
This expression, however, is not gauge invariant, but one can assume that higher-order terms in $J^a$ might turn the ordinary laplacian in (\ref{PsiKKN}) into the adjoint covariant laplacian, and thus convert (\ref{PsiKKN}) again into the form (\ref{PsiGIwithD2}) with  $\mathfrak{K}=\left(\sqrt{-{\cal D}^2+m^2}+m^2\right)^{-1}$.

\begin{figure}[t!]
\fbox{\includegraphics[width=17.5pc]{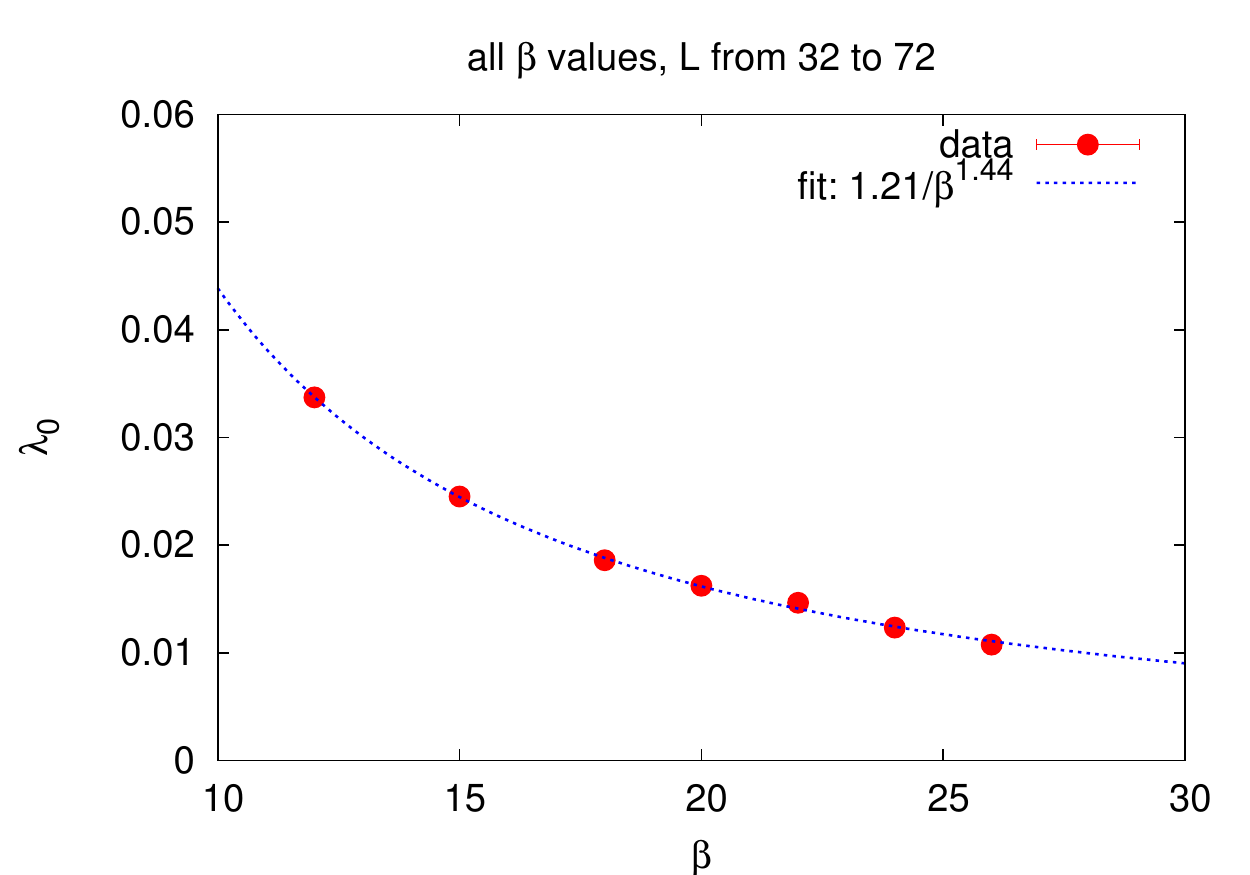}}\hspace{1.5pc}%
\begin{minipage}[b]{18pc}\caption{\label{lambda0}$\lambda_0$ vs.\ $\beta$ from simulations of 3D euclidean SU(2) lattice gauge theory at various couplings $\beta$ and lattice volumes $L^3$. The best fit to data is $\lambda_0\propto\beta^{-1.44},$ which differs from the expected $\beta^{-2}$ dependence and indicates that $\lambda_0$ diverges in the continuum limit.}
\end{minipage}
\end{figure}

\paragraph{\textbf{The proposal of Samuel}}
Almost 20 years ago, Samuel \cite{Samuel:1996bt} put forward a simple vacuum wave functional of the form (\ref{PsiGIwithD2}), interpolating between weak-coupling (\ref{PsiQED}) and DR (\ref{PsiDR}) limits:
\begin{equation}\label{PsiSamuel}
\mathbf{\Psi}_0[A]=
{\cal{N}}\exp\left[-{\textstyle\frac{1}{2}}\int d^2x\;d^2y\; 
B^a(x)\;
\left(\frac{1}{\sqrt{{-{\cal D}^2+m_0^2}}}\right)_{xy}^{ab}
\;B^b(y)\right].
\end{equation}
He estimated with its use the 0${}^{++}$ glueball mass to be about 1.5 GeV. However, this particular form may be flawed if taken at face value: there are hints that the adjoint covariant laplacian needs to be regularized. We have computed its eigenvalues in numerical simulations of the three-dimensional euclidean SU(2) YM theory on a lattice. $(-{\cal D}^2)$ has a positive definite spect\-rum, finite with a lattice regulator. If its lowest eigenvalue $\lambda_0$ were finite in the continuum limit, it should scale, as a function of $\beta=4/g^2$, as $\beta^{-2}$ for large $\beta$. Our data (see figure~\ref{lambda0}) indicate that $\lim_{\beta\to\infty}\beta^2\lambda_0(\beta)\to\infty$, \textit{i.e.\/} $\lambda_0$
diverges for typical configurations in the continuum limit.

%\begin{center}
%\end{center}

\paragraph{\textbf{The GO proposal}}
We proposed \cite{Greensite:2007ij} another simple form with a seemingly small, but crucial difference from that of Samuel:
\begin{equation}\label{PsiGO}
\mathbf{\Psi}_0[A]=
{\cal{N}}\exp\left[-{\textstyle\frac{1}{2}}\int d^2x\;d^2y\; 
B^a(x)\;
\left(\frac{1}{\sqrt{{(-{\cal D}^2-\lambda_0)+m^2}}}\right)_{xy}^{ab}
\;B^b(y)\right].
\end{equation}
As a remedy to the problem mentioned above, it is suggested here to subtract from the covariant laplacian its lowest eigenvalue. The proposed VWF contains a single free parameter, a~mass~$m$, that should vanish in the free-field limit (for $g\to0$). The expression is assumed to be regularized on a lattice, and we use the simplest discretized form of the adjoint covariant laplacian:
\begin{equation}\label{D2lattice}
\left({-{\cal D}^2}\right)^{ab}_{xy}=
4\;\delta^{ab}\delta_{xy}-\sum_{k=1}^2 \left[{\cal U}^{ab}_k(x)\;\delta_{y,x+\hat{k}}+{\cal U}^{\dagger ba}_k(x-\hat{k})\;\delta_{y,x-\hat{k}}\right],
\end{equation}
where 
\begin{equation}\label{Uadj}
{\cal U}^{ab}_k(x)=\frac{1}{2}\mbox{Tr}\left[\sigma^a U_k(x) \sigma^b U^\dagger_k(x)\right]
\end{equation}
and $U_k(x)$ are the link matrices in the fundamental representation.

\begin{figure}[t!]
\begin{minipage}[t]{18pc}
\fbox{\includegraphics[width=17.5pc]{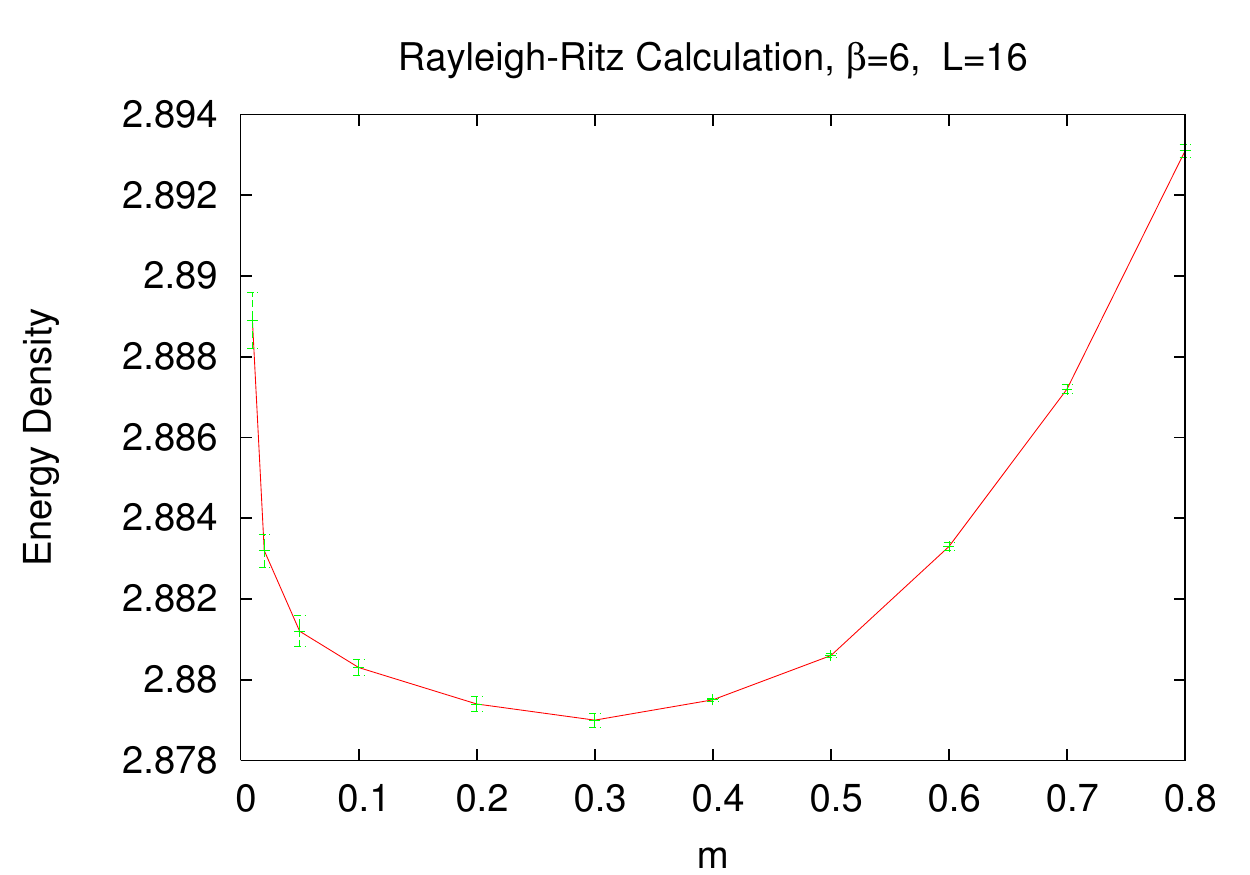}}
\caption{\label{ritz6L16}Vacuum energy density vs. mass parameter $m$ from nu\-me\-ri\-cal simulations on a lattice with $L=16$ at coupling $\beta=6$. The minimum is away from zero, at roughly $m=0.3$. This gives a string tension which
is a little low for $\beta=6$, but the disagreement should not be taken too seriously, because
the estimate for vacuum energy (\ref{variational}) is only approximate.}
\end{minipage}
\hspace{1.5pc}%
\begin{minipage}[t]{18pc}
\fbox{\includegraphics[width=17.5pc]{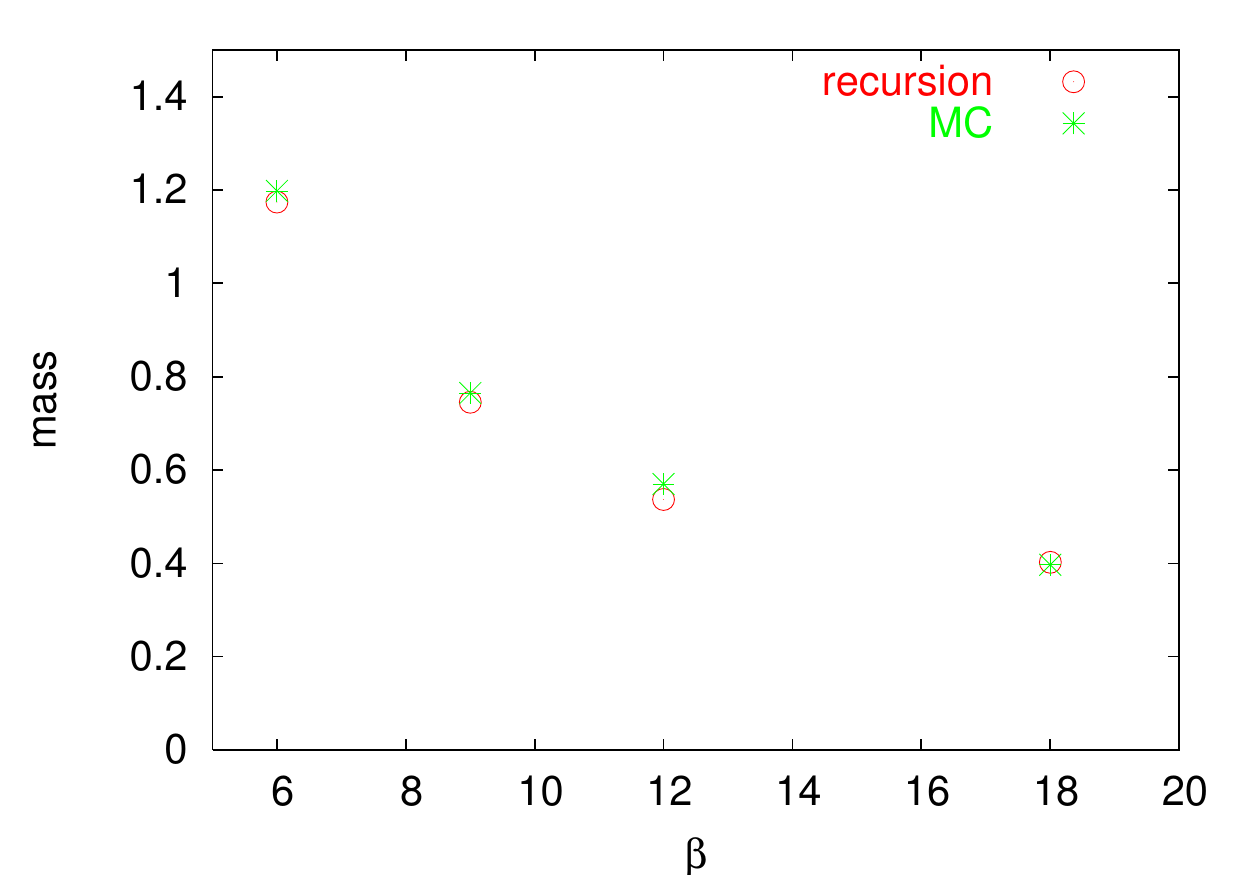}}
\caption{\label{gap}Mass gaps, extracted from equal-time correlators of colour magnetic fields \cite{Greensite:2007ij}, measured in recursion lattices at various lattice couplings, compared to the $0^{++}$ glueball masses in $(2+1)$ dimensions obtained via standard
lattice Monte Carlo methods (from \cite{Meyer:2003wx}, denoted ``MC''). They agree within a few ($<6$) per cent.}
\end{minipage}
\end{figure}

\medskip
We have provided a number of (semi)analytic arguments in favour of the proposed VWF~(\ref{PsiGO}):
\begin{enumerate}
\item
$\mathbf\Psi_0$ reproduces the VWF of electrodynamics (\ref{PsiQED}) in the free-field limit (for $g\to0$).
\item
The proposed form is a good approximation to the true vacuum also for strong fields constant in space and varying
only in time \cite{Greensite:2007ij}. Indeed, if we put such a physical system into a finite volume $V$, its lagrangian and hamiltonian are:
\begin{equation}\label{LHstrong2d}
{{\cal L}={\displaystyle\frac{1}{2}}V
\left(\displaystyle\sum_{k=1}^2\partial_t\vec{A_k}\cdot\partial_t\vec{A_k}
-g^2 \mathfrak{S}^2\right)},
\qquad
{\hat{\cal H}=-\displaystyle\frac{1}{2V}
\sum_{k=1}^2\frac{\partial^2}{\partial\vec{A_k}\cdot\partial\vec{A_k}}
+\frac{1}{2}g^2V \mathfrak{S}^2},
\end{equation}
where ${\mathfrak{S}}=\vert\vec{A}_1\times\vec{A}_2\vert$. It is natural to look for the ground-state solution of the Schr\"odinger equation $\hat{\cal H}\mathbf{\Psi}=E\mathbf{\Psi}$ in the form of $1/V$ expansion:
\begin{equation}\label{1overV}
\mathbf{\Psi}_0=\exp[-VR_0-R_1-V^{-1}R_2-\dots].
\end{equation}
The leading term has to satisfy
\begin{equation}\label{leading}
V\left[-\sum_{k=1}^2\frac{\partial R_0}{\partial\vec{A_k}}\cdot\frac{\partial R_0}{\partial\vec{A_k}}
+g^2\mathfrak{S}^2\right]
=0 \quad [\ +\ \mbox{terms of }{\cal{O}}(1/V)],
\end{equation}
and is easily found to be:
\begin{equation}\label{R0}
R_{0} = \frac{1}{2}g\frac{\mathfrak{S}^2}{\mathfrak{L}}, 
\end{equation}
where $\mathfrak{L}={\sqrt{\vec{A}_1\cdot\vec{A}_1+\vec{A}_2\cdot\vec{A}_2}}$.
The colour vectors $\vec{A}_1$ and $\vec{A}_2$ define a plane in colour space, we can choose \textit{e.g.\/} $\vec{A}_1=({\cal A}_1,0,0), \vec{A}_2=({\cal A}_2\cos\theta,{\cal A}_2\sin\theta,0)$, then
${\mathfrak{S}}=\vert{\cal A}_1{\cal A}_2\sin\theta\vert$. We assume ${\cal{A}}_1,{\cal{A}}_2$ to be of the same order ${\cal{O}}({\cal{A}})$. The contribution of $R_0$ to (\ref{1overV}) will be non-negligible if $VR_0\sim{\cal O}(1)$, \textit{i.e.\/} $R_0\sim{\cal O}(1/V)$.

On the other hand, for strong enough fields, ${\cal{D}}_k^{ac}\approx g\epsilon^{abc}A_k^{b}$ and
\begin{equation}\label{D2strong}
\left(-{\cal{D}}^2\right)^{ab}_{xy}\approx g^2\delta_2(x-y)
\left[(\vec{A}_1^2+\vec{A}_2^2)\delta^{ab}-A_1^aA_1^b-A_2^aA_2^b\right]
\equiv g^2 \delta_2(x-y)\;{\cal Z}^{ab},
\end{equation}
where
\begin{equation}\label{Z}
{\cal Z}=\left(\begin{array}{c c c}
{\cal A}_2^2\sin^2\theta & -{\cal A}_2^2\sin\theta\cos\theta & 0\\
-{\cal A}_2^2\sin\theta\cos\theta & {\cal A}_1^2+{\cal A}_2^2\cos^2\theta & 0\\
0 & 0 &  {\cal A}_1^2+{\cal A}_2^2
\end{array}\right).
\end{equation}
The leading term in our VWF (\ref{PsiGO}) then is
\begin{equation}\label{PsiGOstrong2}
\mathbf{\Psi}_0={\cal{N}}\exp\left[-\frac{1}{2}gV \frac{\mathfrak{S}^2}{{\sqrt{\zeta_3-\zeta_1+(m/g)^2}}}\right],
\end{equation}
where $\zeta_1$ and $\zeta_3$ are respectively the smallest and largest eigenvalues of the above matrix~$\cal{Z}$. For strong constant fields ($\vert g{\cal{A}}\vert\gg m,\lambda_0$), both $(m/g)^2$ and $\zeta_1\sim R_0\sim{\cal{O}}(1/V)$ are negligible w.r.t.\ $\zeta_3=\mathfrak{L}^2$, and the expression (\ref{PsiGOstrong2}) agrees with (\ref{1overV}) and (\ref{R0}).
\item
If we split the colour magnetic field strength $\mathbf{B}(x)$ into ``fast'' and ``slow'' components, the part of the VWF that depends on
$\mathbf{B}_\mathrm{slow}$ reduces to the magnetic-disorder (DR) form (\ref{PsiDR}). The fundamental string tension is then easily computed as $\sigma_F=3mg^2/16=3m/4\beta$. Non-zero value of the parameter $m$ then implies non-zero $\sigma_F$, \textit{i.e.\/} confinement of fundamental-representation colour charges.
\item
One can take the mass $m$ in the wave functional as a free variational parameter and compute (approximately) the
expectation value of the YM hamiltonian (see section VI of \cite{Greensite:2007ij} for details). The result, expressed as a sum over eigenvalues of the adjoint covariant laplacian in the thermalized gauge field $\mathbf{A}$, is:
\begin{equation}\label{variational}
\langle \hat{\cal H}\rangle
=\frac{1}{2}
\left\langle{\displaystyle\sum_n}\left(
\sqrt{\lambda_n-\lambda_0+m^2}+
\frac{1}{2}\displaystyle\frac{\lambda_0-m^2}
{\sqrt{\lambda_n-\lambda_0+m^2}}\right)\right\rangle.
\end{equation}
In the abelian free-field case, the optimal value of $m$ equals $\lambda_0$ and $\lambda_0$ goes to 0 in the continuum limit, so the theory is non-confining. In the non-abelian case one can evaluate the energy density $\langle \hat{\cal H}\rangle/L^2$ numerically in Monte Carlo simulations, and finds that a non-zero (finite) value of $m$ is energetically preferred. A typical result is displayed in figure~\ref{ritz6L16}. 
\end{enumerate}

%\begin{center}
%\end{center}

\paragraph{\textbf{Some numerical evidence}}
All above arguments are encouraging, but quantitative tests are needed to gain more confidence in the proposed form of the VWF. Such tests are provided by numerical lattice simulations. For readers not familiar with the method, basics is summarized in \ref{A}.

We have compared a set of physical quantities computed in two ensembles of lattice gauge-field configurations:
\begin{enumerate}
\item[I.]%
{\textit{Monte Carlo lattices\/}: Ensemble of two-dimensional slices of configurations generated by Monte Carlo  simulations of three-dimensional euclidean SU(2) lattice gauge theory with standard Wilson action (\ref{S_W}) at a coupling $\beta=4/g^2$; from each configuration, only one (random) slice at fixed euclidean time was taken. These configurations are distributed with the weight proportional to the square of the true VWF of the theory, $\vert\mathbf{\Psi}_\mathrm{true}[U]\vert^2$.}
\item[II.]%
{\textit{``Recursion'' lattices\/}: Ensemble of independent two-dimensional lattice configurations generated with the probability distribution given by the (square of the) VWF (\ref{PsiGO}), with $m$ and $g^2$ fixed to get the correct value of the fundamental string tension. These configurations can be generated efficiently by the recursion method proposed (and described in detail) in~\cite{Greensite:2007ij}. Essential points of the method are sketched in \ref{B}.}
\end{enumerate}

\begin{figure}[t!]
\begin{minipage}[t]{18pc}
\fbox{\includegraphics[width=17.5pc]{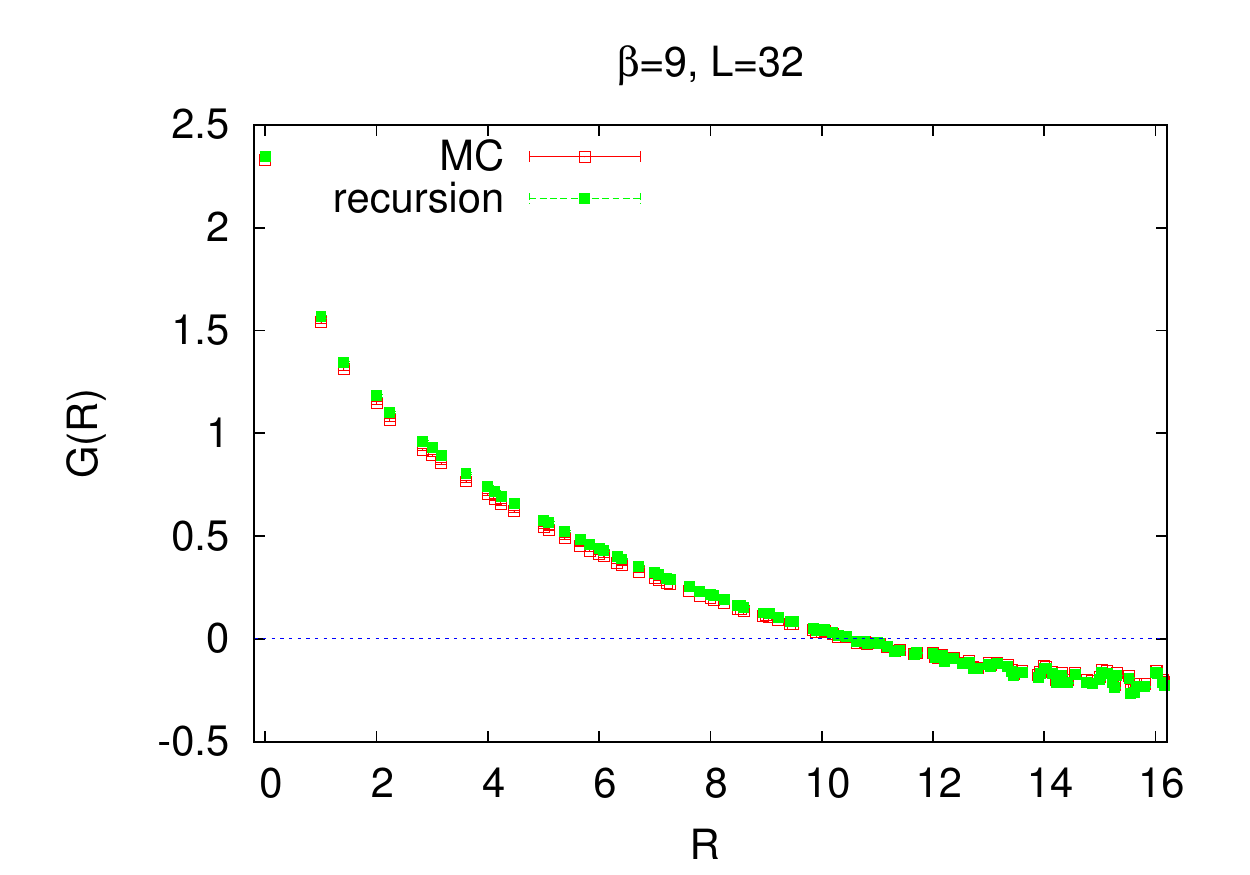}}
\caption{\label{propagator_b9_l32_all}The Coulomb-gauge ghost pro\-pagator  at $\beta=9$ on $32^2$ lattice.}
\end{minipage}\hspace{1.5pc}%
\begin{minipage}[t]{18pc}
\fbox{\includegraphics[width=17.5pc]{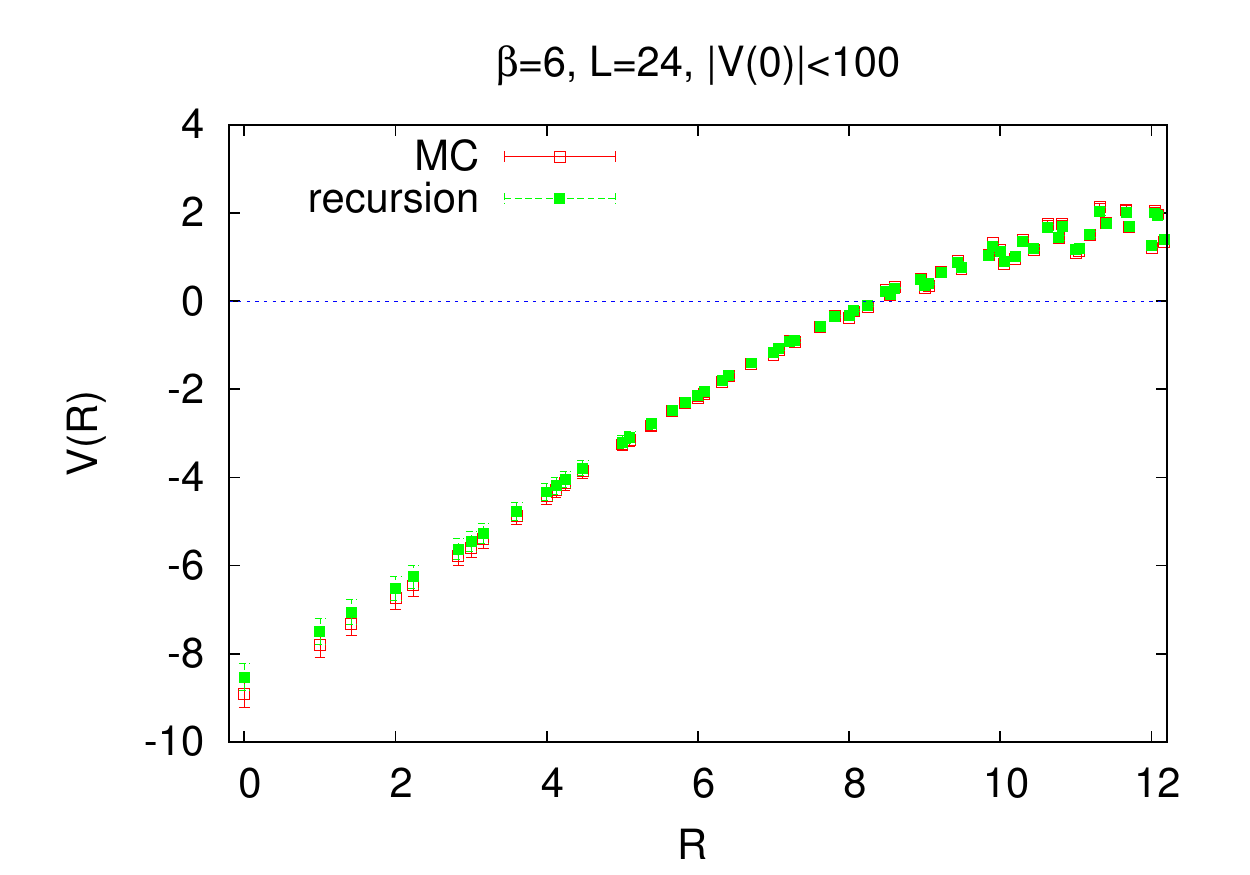}}
\caption{\label{potential_b6_l24_l100}The colour-Coulomb potential at $\beta=6$ on $24^2$ lattice, computed from con\-figurations with a cut $\vert V(0)\vert<100$.}
\end{minipage} 
\end{figure}

We computed in both ensembles the mass gap \cite{Greensite:2007ij}, the Coulomb-gauge ghost propagator, and the colour Coulomb potential \cite{Greensite:2010tm}. The latter two quantities are of particular interest because of their role in the so-called Gribov--Zwanziger mechanism of confinement~\cite{Gribov:1977wm,Zwanziger:1998ez}. Example results are shown in figures  \ref{gap}, \ref{propagator_b9_l32_all}, and \ref{potential_b6_l24_l100}. The agreement between quantities measured  in MC and recursion ensembles is very reasonable.\footnote{The agreement in the case of the colour Coulomb potential is less satisfactory and we attribute it to the existence of exceptional configurations that are extremely difficult to fix to the Coulomb gauge. Differences in the measured colour Coulomb potentials in two ensembles are small, if one ensures approximately equal population of exceptional configurations in both of them. (See Section IV of \cite{Greensite:2010tm} for details on this point.)}

%\begin{center}
%\end{center}

One can also measure (\textit{i.e.\/} compute in numerical simulations) amplitudes of various sets of test con\-figurations in the true YM vacuum, and compare them with predictions based on the proposed VWF (\ref{PsiGO}). The method for computing relative weights of configurations was proposed by Greensite and Iwasaki long ago \cite{Greensite:1988rr} and will be described in section \ref{spaceland}. The
technique was applied in $(2+1)$ dimensions to sets of abelian plane wave configurations of varying amplitude and wavelength, and sets of non-abelian constant configurations \cite{Greensite:2011pj}. The obtained data agree with expectations based on (\ref{PsiGO}), but similar agreement was obtained also for some other forms of wave functionals which simplify to a DR form at large scales (see \cite{Greensite:2011pj}).

Summarizing this section, we found analytic and numerical evidence that in $D=2+1$ dimensions our Ansatz for the VWF seems a fairly good approximation to the true ground state of the theory. However, we do not live in \textit{Flatland\/}, our world is \textit{Spaceland\/}, so I will switch for the rest of this contribution to pure Yang--Mills theory in four spacetime dimensions. 

\section{Version in 3D or Spaceland%
\footnote{Not surprisingly, one can find also a book with \textit{Spaceland\/} in the title \cite{Rucker:2002aa}. It is a science-fiction novel  written by American mathematician and computer scientist Rudy Rucker as a tribute to Edwin Abbott's \textit{Flatland\/}.}}\label{spaceland}
The extension of the proposed vacuum wave functional (\ref{PsiGO}) to $(3 + 1)$ dimensions is straight\-forward: one replaces
the product $B^a(x)\;B^b(y)$ by $\frac{1}{2}F^a_{ij}(x)\;F^b_{ij}(y)$, and two-dimensional integrals by three-dimensional ones:
\begin{equation}\label{PsiGO3D}
\mathbf{\Psi}_0[A]=
{\cal{N}}\exp\left[-{\textstyle\frac{1}{4}}\int d^3x\;d^3y\; 
F^a_{ij}(x)\;
\left(\frac{1}{\sqrt{{(-{\cal D}^2-\lambda_0)+m^2}}}\right)_{xy}^{ab}
\;F^b_{ij}(y)\right].
\end{equation}
Of course, the adjoint covariant laplacian now involves $k$-summation over three space directions:
\begin{equation}\label{D2lattice3D}
\left({-{\cal D}^2}\right)^{ab}_{xy}=
6\;\delta^{ab}\delta_{xy}-\sum_{k=1}^3 \left[{\cal U}^{ab}_k(x)\;\delta_{y,x+\hat{k}}+{\cal U}^{\dagger ba}_k(x-\hat{k})\;\delta_{y,x-\hat{k}}\right].
\end{equation}

\paragraph{\textbf{Positive news, but\dots}}
The wave functional (\ref{PsiGO3D}) is also in $(3+1)$ dimensions exact in the free-field limit. It also approximately solves the YM Schr\"odinger equation in the zero-mode, strong-field limit. It is now convenient to introduce:
\begin{eqnarray}
{\mathfrak{L}}&=&\sqrt{\vec{A}_1\cdot \vec{A}_1+\vec{A}_2\cdot\vec{A}_2+\vec{A}_3\cdot\vec{A}_3},\\
{\mathfrak{S}}&=&\sqrt{(\vec{A}_1\times \vec{A}_2)^2+(\vec{A}_2\times \vec{A}_3)^2+(\vec{A}_3\times \vec{A}_1)^2},\\
{\mathfrak{V}}&=&\left\vert\vec{A}_1\cdot(\vec{A}_2\times \vec{A}_3)\right\vert.
\end{eqnarray}
Then the leading term $R_0$ of the $1/V$ expansion (\ref{1overV}) of the exponent of $\mathbf{\Psi}_0$ is again given by~(\ref{R0}) since
\begin{equation}\label{leading3D}
V\left[-\sum_{k=1}^2\frac{\partial R_0}{\partial\vec{A_k}}\cdot\frac{\partial R_0}{\partial\vec{A_k}}
+g^2\mathfrak{S}^2\right]=0+{g^2 V\left(\frac{7{\mathfrak{S}}^4}{4\mathfrak{L}^4}-\frac{3{\mathfrak{V}}^2}{\mathfrak{L}^2}\right)}
=0 + {\cal{O}}\left(\frac{1}{V}\right)
\end{equation}
for strong fields from the \textit{``abelian valley''\/} where the components $\vec{A}_1, \vec{A}_2, \vec{A}_3$ are
nearly aligned, or antialigned, in colour space.

In this same limit, the proposed VWF (\ref{PsiGO3D}) reduces to $\mathbf{\Psi}_0={\cal{N}}\exp(-Q)$ with 
\begin{equation}\label{Q}
Q\approx{\textstyle\frac{1}{4}} gV (\vec{A}_{i}\times\vec{A}_{j})^{a} \left( {\delta^{ab}\over {\mathfrak{L}} }
 - {\delta^{a3}\delta^{b3}\over{\mathfrak{L}}} + {\delta^{a3}\delta^{b3}\over m} \right)
                (\vec{A}_{i}\times\vec{A}_{j})^{b}.
\end{equation}
The first term dominates and gives $VR_0\sim{\cal{O}}(1)$ as above, the other two are of order ${\cal{O}}\left(1/V\right)$  (see Appendix A of \cite{Greensite:2007ij}).

However, the transition to three space dimensions causes complications associated with the Bianchi constraint that colour magnetic fields $F^a_{ij}$ have to satisfy. Because of that constraint, direct numerical generation of gauge-field configurations with the probability distribution given by the square of the vacuum wave functional (\ref{PsiGO3D})  is much more challenging. I have not yet been able to find a way of generating such configurations, similar to the recursion method used in two space dimensions (see section \ref{flatland} and \ref{B}). A different approach to testing our proposal is required.

\begin{figure}[t!]
\fbox{\includegraphics[width=17.5pc]{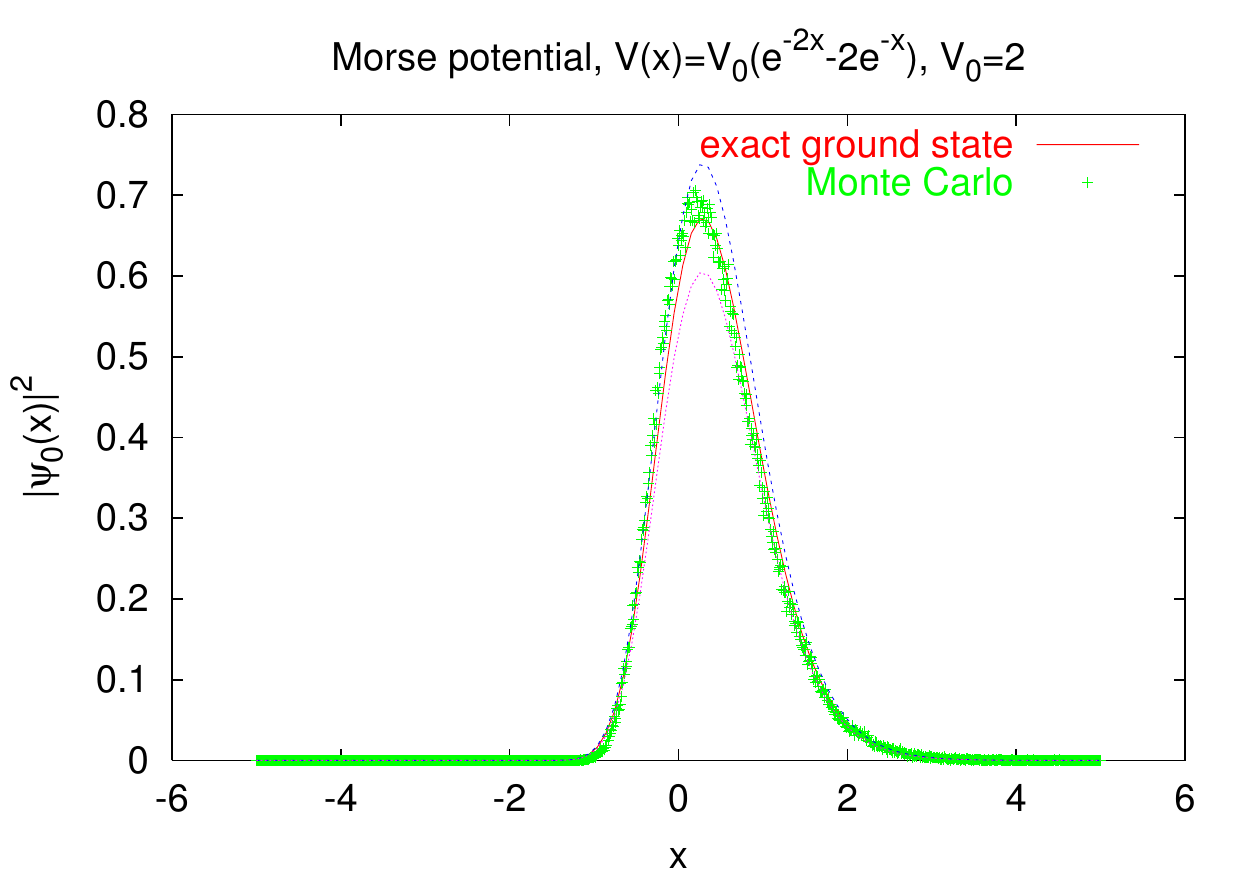}}\hspace{1.5pc}%
\begin{minipage}[b]{18pc}\caption{\label{morse}Result of a numerical determination of $\vert\psi_0(x)\vert^2$ for the Morse potential $V(x)=V_0\left(e^{-2x}-2e^{-x}\right)$ with $V_0=2$ \cite{Brndiar:2002aa}. The red solid line shows the exact result, the green points are the outcome of a Monte Carlo computation based on (\ref{psi2path}), and dotted lines indicate the $\pm 10$\% band around the exact solution.}
\end{minipage}
\end{figure}

\paragraph{\textbf{Measurement of the ground-state wave function in quantum mechanics}}
Let us start with a simple question: How can one numerically compute the wave function of a ground-state of a one-dimensional system with hamiltonian $H$ in quantum mechanics? The most direct approach is to solve the Schr\"odinger equation numerically. A less accurate way, but useful for our purposes, is to start from a simple formula:
\begin{equation}\label{psi2QM}
\vert\psi_0(x)\vert^2=\lim_{\tau\to\infty}e^{E_0 \tau}G(x,-i\tau;x,0)=\lim_{\tau\to\infty}\frac{G(x,-i\tau;x,0)}{\int d\xi\;G(\xi,-i\tau;\xi,0)},
\end{equation}
where $G(x_2,-i\tau;x_1,0)$ is the Green's function (propagator) of a free particle moving from the point $x_1$ at $t=0$ to the point $x_2$ at euclidean time $(-i\tau)$. Expressing the ratio in (\ref{psi2QM}) through path integrals in the usual way (see \textit{ e.g.\/} \cite{Wipf:2013vp})
\begin{equation}\label{Gpath}
\frac{G(x_N=x,-i\tau;x_0=x,0)}{\int dx\;G(x_N=x,-i\tau;x_0=x,0)}=
\frac{\int dx_1\dots dx_{N-1}\exp\left[-\int_0^\tau Hd\tau'\right]}{\int dx_1\dots dx_{N-1}dx_N\exp\left[-\int_0^\tau Hd\tau'\right]},
\end{equation}
one finally gets
\begin{equation}\label{psi2path}
\vert\psi_0(x)\vert^2=\displaystyle\frac{1}{Z}\int[D\xi(t)]\;\delta\left[\xi(0)-x\right]\;\e^{-S\left[\xi,\dot{\xi}\right]}\;,
\end{equation}
where $S$ is the euclidean action corresponding to the hamiltonian $H$. This equation expresses the wave function squared as an average of a $\delta$-function over all paths, a procedure that might appear totally inappropriate for numerical computation. However, with a simple trick it \textit{can\/} be implemented efficiently \cite{Landau:2008aa}, and a sample result is displayed in figure \ref{morse}.

%\begin{center}
%\end{center}

\paragraph{\textbf{The relative-weight method}}
The method of Greensite and Iwasaki \cite{Greensite:1988rr} is based on a generalization of (\ref{psi2path}) to quantum field theory. The squared VWF of the pure YM theory is given by the path integral\footnote{One could insert into the path integral a factor imposing fixing to lattice temporal gauge, which sets all timelike links to $\mathbf{1}$ except on one time slice at $t\ne 0$. However, this gauge fixing is in fact not necessary.}:
\begin{equation}\label{psi2YM}
\mathbf{\Psi}^2_0[U']=\displaystyle\frac{1}{Z}\int [DU]\;\prod_{\mathbf{x},i}\delta[U_i(\mathbf{x},0)-U'(\mathbf{x})]\;\exp\left(-S[U]\right),
\end{equation}
The relative-weight method  \cite{Greensite:1988rr} enables one to compute ratios $\mathbf{\Psi}_0^2[U^{(n)}]/\mathbf{\Psi}_0^2[U^{(m)}]$  for configurations belonging to a finite set {${\mathcal{U}}=\left\lbrace U_i^{(j)}(\mathbf{x}),j=1,2,\dots,M\right\rbrace$} using a simple procedure: One performs Monte Carlo simulations with the usual update algorithm (\textit{e.g.\/}\ heat-bath) for all spacelike links at $t\ne0$ and for timelike links. Once in a while one updates the spacelike links at $t=0$ all at once selecting one configuration from the set $\mathcal{U}$ at random, and accepts/rejects it via the Metropolis prescription. Then, for a large number of updates $N_\mathrm{tot}$,
\begin{equation}\label{psi2ratio}
\frac{\mathbf{\Psi}_0^2[U^{(n)}]}{\mathbf{\Psi}_0^2[U^{(m)}]}=\lim_{N_\mathrm{tot}\to\infty}\frac{N_n}{N_m},
\end{equation}
where $N_m$ ($N_n$) denotes the number of times the $m$-th ($n$-th) configuration is accepted. To~ensure a non-negligible acceptance rate of Metropolis updates for all configurations in the set $\mathcal{U}$, they must lie close in configuration space. This limits somewhat the applicability of the method.

\begin{figure}[t!]
\begin{minipage}[b]{18pc}
\fbox{\includegraphics[width=17.5pc]{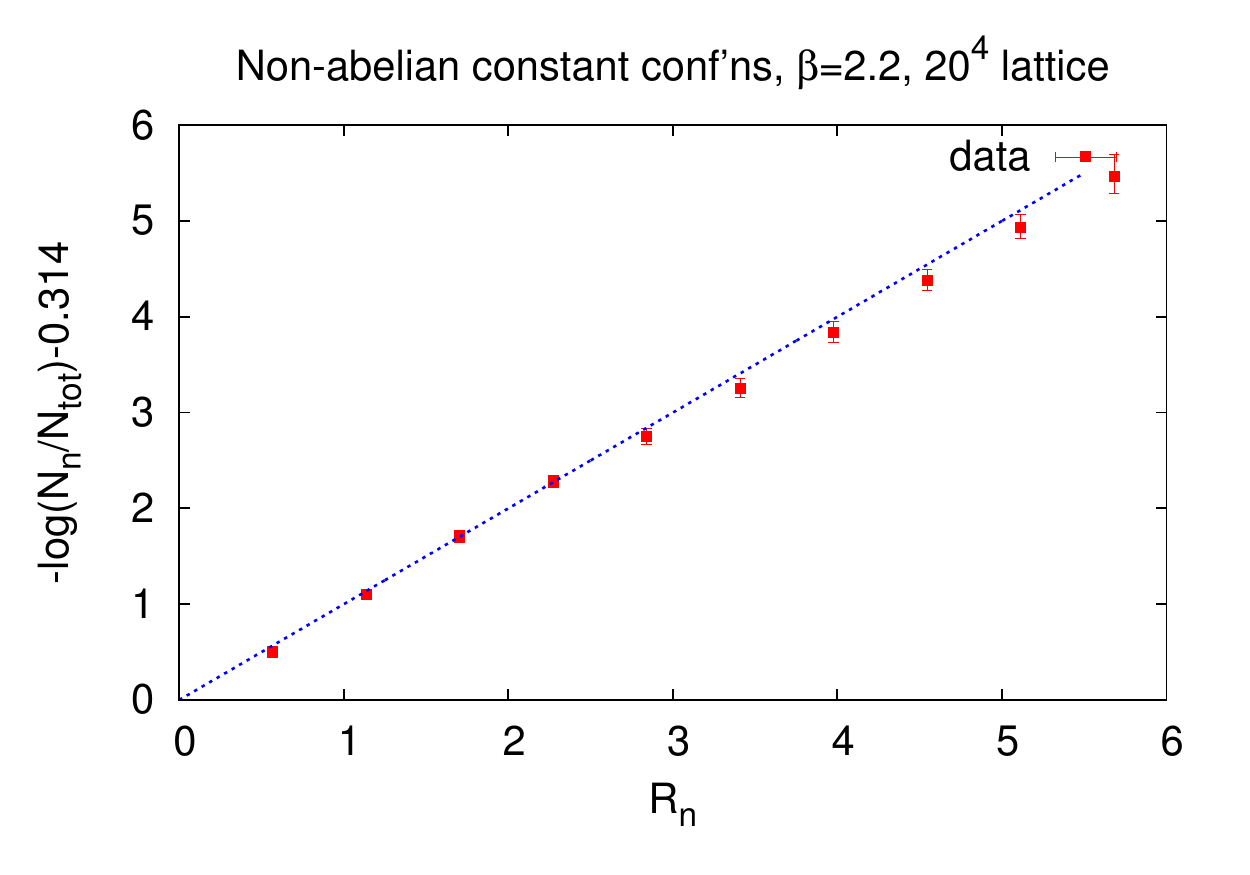}}
\caption{\label{prob_k_2_2_l20_c}$[-\log(N_n/N_\mathrm{tot})]$ vs.\ $R_n=\mu\kappa n$ for non-abelian constant configurations with $\kappa=0.14$; $\mu$ at $\beta=2.2$ on $20^4$ lattice was $4.06 (4)$.}
\end{minipage}
\hspace{1.5pc}%
\begin{minipage}[b]{18pc}
\fbox{\includegraphics[width=17.5pc]{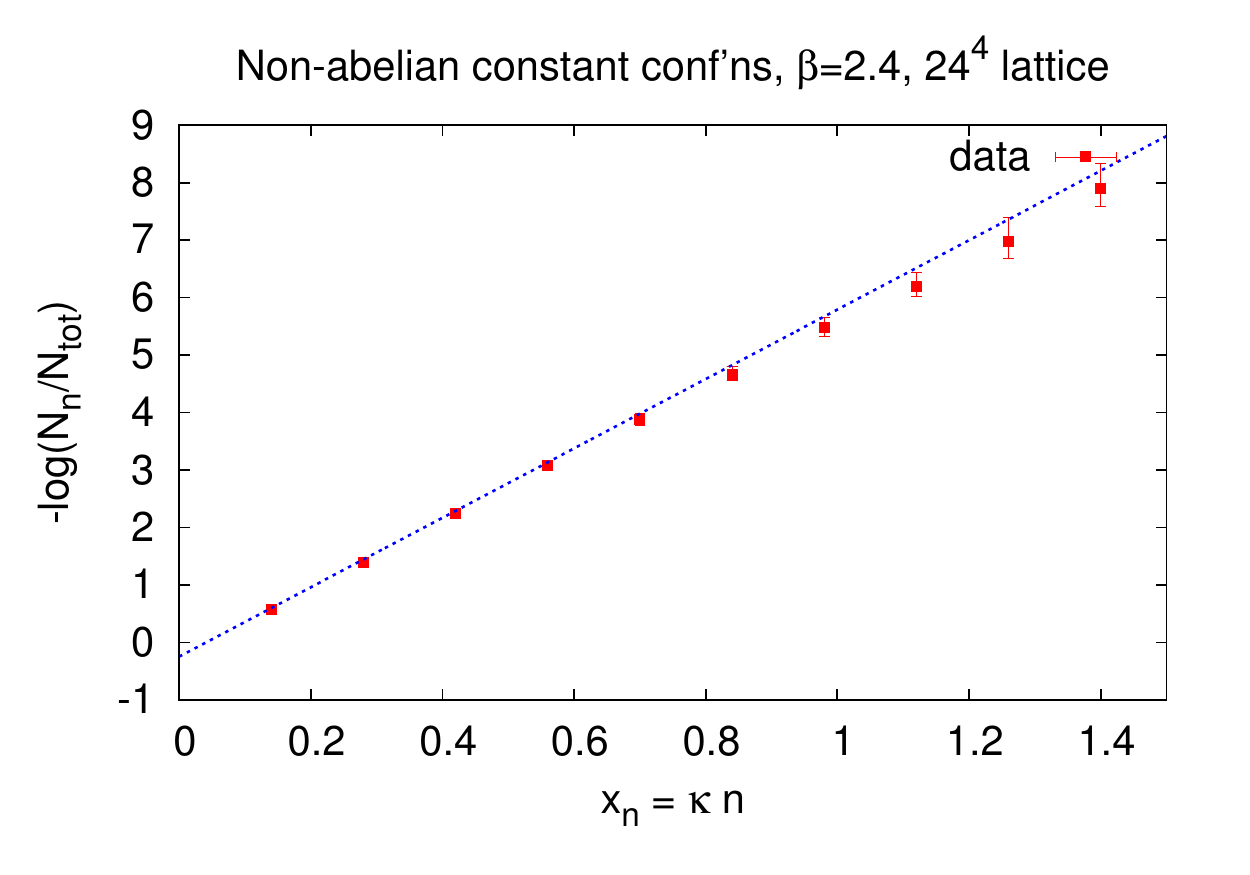}}
\caption{\label{prob_k_2_4_l24_n}$[-\log(N_n/N_\mathrm{tot})]$ vs.\ $x_n=\kappa n$ for NAC configurations with $\kappa=0.14$ at $\beta=2.4$ on $24^4$ lattice; the slope $\mu$ from a linear fit comes out $6.04 (2)$.}
\end{minipage}
\end{figure}

If the VWF is assumed to be of the form $\mathbf\Psi_0[U]={\cal{N}}\exp(-R[U])$, then the measured values of $[-\log(N_n/N_\mathrm{tot})]$ should fall on a straight line with unit slope as function of $R_n\equiv R[U^{(n)}]$. An example is shown in figure \ref{prob_k_2_2_l20_c} for a set of non-abelian constant configurations (to be specified below).

%\begin{center}
%\end{center}

\paragraph{\textbf{Direct measurement of the vacuum wave functional}} 
We used the relative-weight method to calculate the squared WVF for the following two classes of lattice gauge-field configurations:
\begin{itemize}
\item[I.]
\textit{Non-abelian constant (NAC) configurations\/}:
\begin{equation}\label{U_NAC}
{\cal U}_\mathrm{NAC}=\left\{U_k^{(n)}(x)=\sqrt{1-\left(u^{(n)}\right)^2}\;\mathbf{1}+iu^{(n)}\bm{\sigma}_k\right\},
\end{equation}
where
\begin{equation}\label{U_NAC_2}
u^{(n)}=\left(\frac{\kappa}{6L^3}n\right)^{1/4},\qquad n\in\{1,2,\dots, 10\}.
\end{equation}
The constant $\kappa$, regulating amplitudes of NAC configurations, is selected so that the ratio of the smallest to the largest weight within the set is not too small, at most ${\cal{O}}\left(10^{-4}\div10^{-3}\right)$.

The expected dependence of relative weights on $\kappa$ and $n$ is linear:
\begin{equation}\label{U_NAC_fit}
-\log\left(\frac{N_{n}}{N_\mathrm{tot}}\right)=R_n+\mbox{const.}=\kappa n\times{\mu}+\mbox{const.}
\end{equation}
The data are consistent with this expectation; the constant $\mu$ at a given coupling $\beta$ is obtained as the slope of a linear fit of $[-\log(N_{n}/N_\mathrm{tot})]$ vs.\ $\kappa n$, see figure \ref{prob_k_2_4_l24_n}. This constant coincides with the parameter $\mu$ that appears in the dimensional-reduction Ansatz for the VWF (\ref{PsiDR}).

Predictions for NAC configurations resulting from the DR and our proposed vacuum wave functional are identical (up to parameter-naming conventions). Moreover, the dependence of $\mu$ on $\beta$ was already computed in the pioneering work of Greensite and Iwasaki~\cite{Greensite:1988rr} more than 25 years ago, albeit at rather small-size lattices, $6^4$ and $8^4$. New results from simulations on $20^4$ and $24^4$ lattices confirm their findings, \textit{cf.\/} figure 2 of \cite{Greensite:1988rr} with our figure~\ref{mu_vs_beta_NAC_c}. The strong-coupling prediction $\mu(\beta)=\beta$ is confirmed at small $\beta$, and in the weak-coupling region $\mu(\beta)$ behaves as a physical quantity with the dimension of inverse mass:
\begin{equation}\label{mu_scaling}
\mu(\beta)f(\beta)=\mu_\mathrm{phys}\approx 0.0269(3),
\end{equation}
where
\begin{equation}\label{f_AF}
f(\beta)=\left(\frac{6\pi^2\beta}{11}\right)^\frac{51}{121}\exp\left(-\frac{3\pi^2\beta}{11}\right).
\end{equation}

%\begin{center}
\begin{figure}[t!]
\begin{minipage}[t]{18pc}
\fbox{\includegraphics[width=17.5pc]{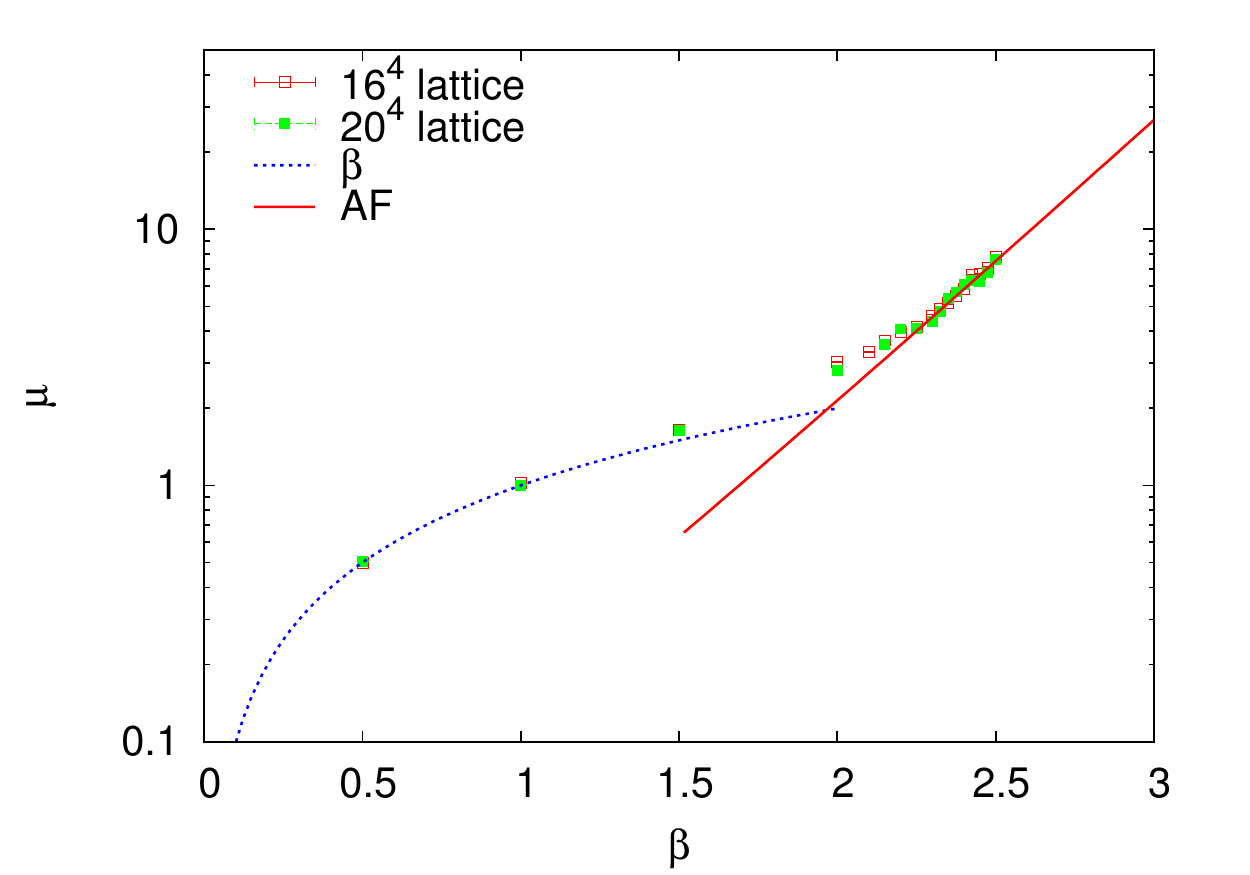}}
\caption{\label{mu_vs_beta_NAC_c}The quantity $\mu$, extracted from computed weights of NAC configurations, vs.\ the coupling $\beta$.}
\end{minipage}
\hspace{1.5pc}%
\begin{minipage}[t]{18pc}
\fbox{\includegraphics[width=17.5pc]{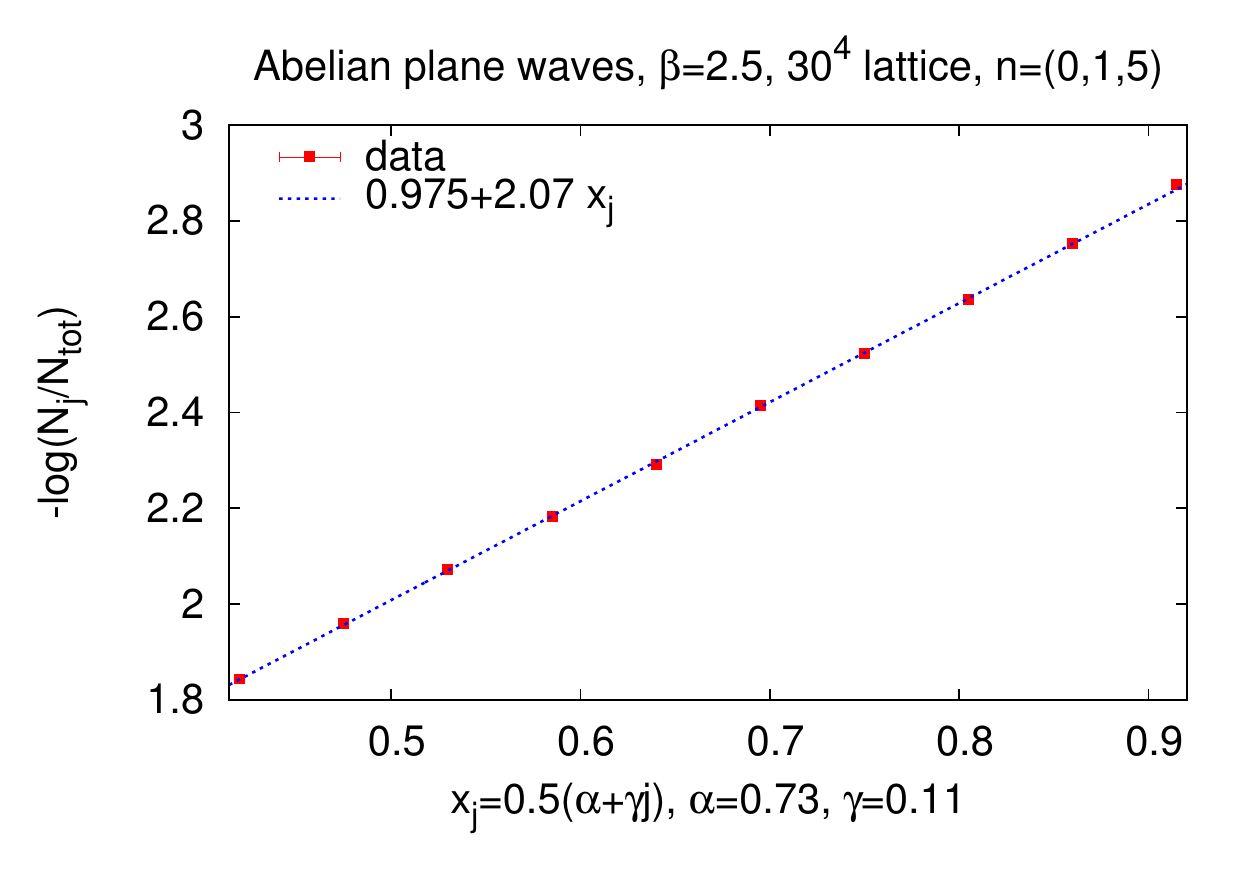}}
\caption{\label{prob_k_a1_2p5_k015_l30_c}An example plot of  $[-\log(N^{(j)}_{\footnotesize\textbf{\textit{n}}}/N_\mathrm{tot})]$ vs.\ ${\textstyle\frac{1}{2}}(\alpha_{\footnotesize\textbf{\textit{n}}}+\gamma_{\footnotesize\textbf{\textit{n}}}j)$ for~APW configurations.}
\end{minipage}
\end{figure}
%\end{center}

%
\item[II.]
\textit{Abelian plane-wave (APW) configurations\/}:
\begin{equation}\label{U_APW}
{\cal U}_\mathrm{APW}=\Biggl\{U_1^{(j)}(x)=\sqrt{1-\left(w^{(j)}_{\footnotesize\textbf{\textit{n}}}(x)\right)^2}\mathbf{1}+iw^{(j)}_{\footnotesize\textbf{\textit{n}}}(x)\bm{\sigma}_3,\qquad
 U_2^{(j)}(x)=U_3^{(j)}(x)=\mathbf{1}\Biggr\},
\end{equation}
where
\begin{equation}\label{U_APW_2}
w^{(j)}_{\footnotesize\textbf{\textit{n}}}=\sqrt{\frac{\alpha_{\footnotesize\textbf{\textit{n}}}+\gamma_{\footnotesize\textbf{\textit{n}}}j}{L^3}}\cos\left(\frac{2\pi}{L}\textbf{\textit{n}}\cdot\textbf{\textit{x}}\right),\qquad \textbf{\textit{n}}=(n_1,n_2,n_3), \qquad j\in\{1,2,\dots, 10\}.
\end{equation}
Amplitudes in a particular set of plane waves with wave number $\textbf{\textit{n}}$ are parametrized by a~pair $(\alpha_{\footnotesize\textbf{\textit{n}}},\gamma_{\footnotesize\textbf{\textit{n}}})$ and depend on integer $j$. For all sets, pairs of parameters were again carefully chosen so that the actions of configurations with different $j$ in the set were not much different.

For APW configurations one expects:
\begin{equation}\label{U_APW_fit}
-\log\left(\frac{N^{(j)}_{\footnotesize\textbf{\textit{n}}}}{N_\mathrm{tot}}\right)=R^{(j)}_{\footnotesize\textbf{\textit{n}}}+\mbox{const.}={\textstyle\frac{1}{2}}(\alpha_{\footnotesize\textbf{\textit{n}}}+\gamma_{\footnotesize\textbf{\textit{n}}}j)\times{\omega(\textbf{\textit{n}})}+\mbox{const.}
\end{equation}
For a particular wave number $\textit{\textbf{n}}$, one can plot $[-\log(N^{(j)}_{\footnotesize\textbf{\textit{n}}}/N_\mathrm{tot})]$ vs.\ ${\textstyle\frac{1}{2}}(\alpha_{\footnotesize\textbf{\textit{n}}}+\gamma_{\footnotesize\textbf{\textit{n}}}j)$, and determine the
slope $\omega(\textbf{\textit{n}})$ from a fit of the form (\ref{U_APW_fit}). The expected linear dependence was seen in all our data sets at all couplings, wave numbers, and parameter choices; an example is displayed in figure \ref{prob_k_a1_2p5_k015_l30_c}.
\end{itemize}

\begin{figure}[t!]
\begin{minipage}[t]{18pc}
\fbox{\includegraphics[width=17.5pc]{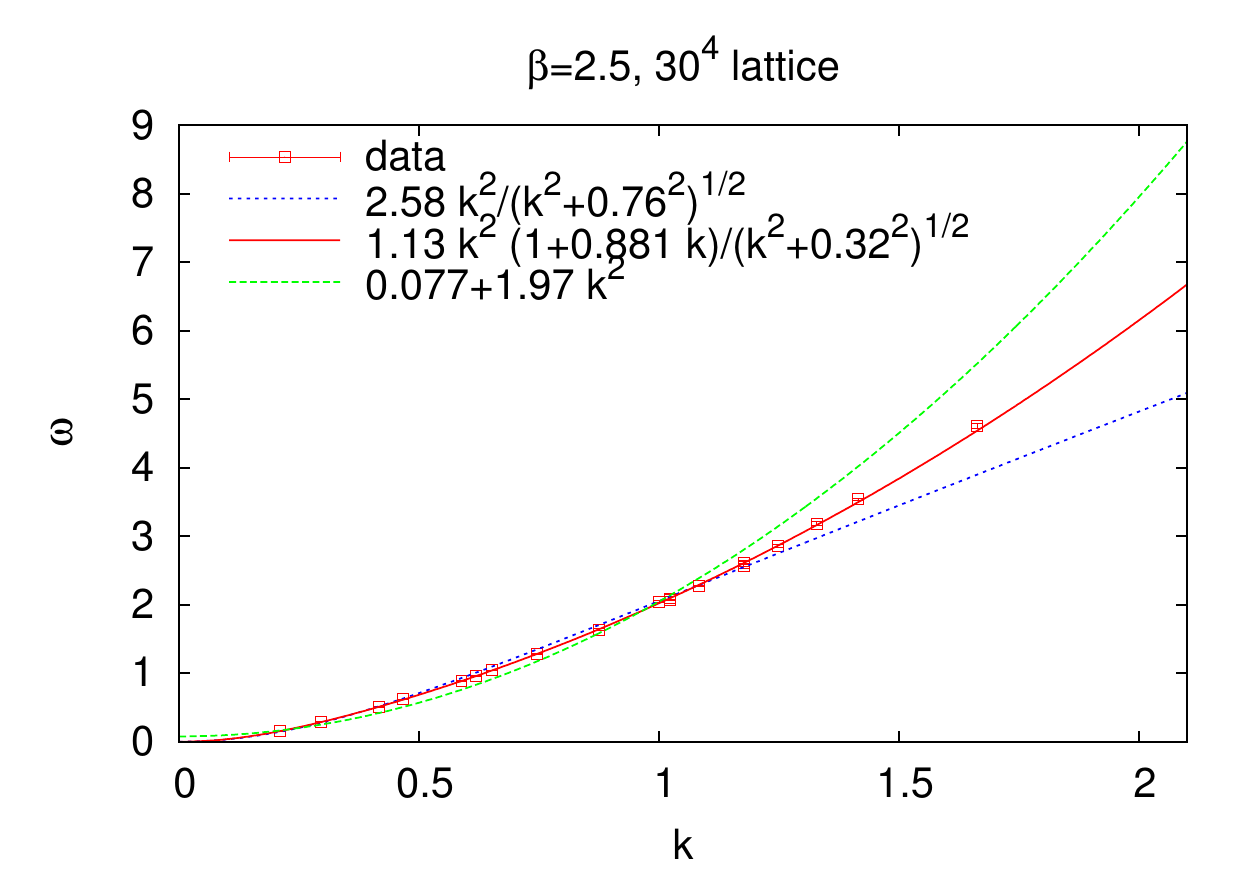}}
\caption{\label{omega_2_5_l30_c}$\omega(\textbf{\textit{n}})$ vs.\ $k(\textbf{\textit{n}})$ with fits of the forms (\ref{fits}) and (\ref{best}) for APW configurations.}
\end{minipage}
\hspace{1.5pc}%
\begin{minipage}[t]{18pc}
\fbox{\includegraphics[width=17.5pc]{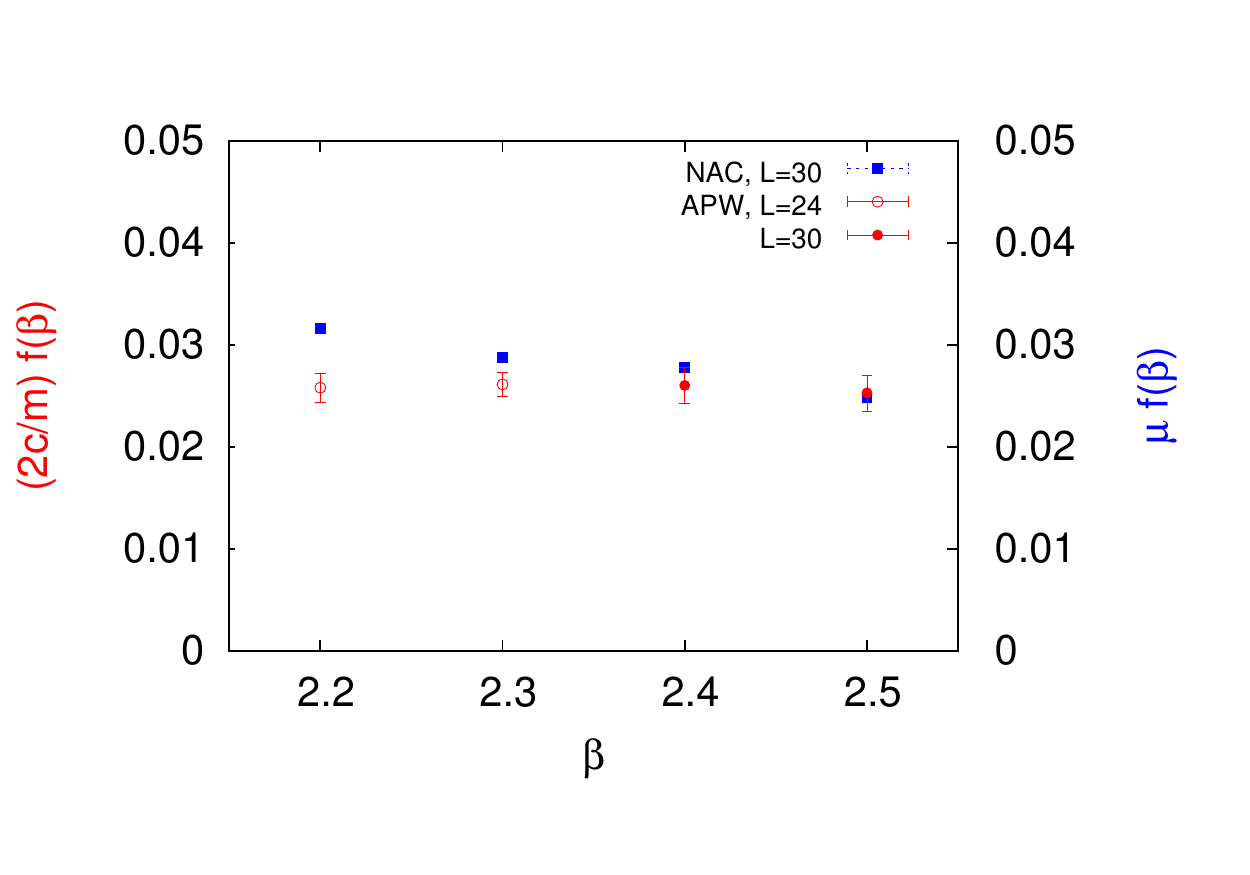}}
\caption{\label{NAC_vs_APW_c}The combination $(2c/m) f (\beta)$ of the best fit (\ref{best}) to APW data. Also displayed are values of $\mu f (\beta)$ extracted from NAC data.}
\end{minipage}
\end{figure}

Our main goal is to compare computed relative weights of test configurations with predictions of the DR (\ref{PsiDR}) and GO (\ref{PsiGO3D}) wave functionals. As already mentioned, NAC configurations are not suitable for that purpose, they only served us as test bed of our computer code. However, for APW the DR prediction for the dependence of the extracted function $\omega(\textbf{\textit{n}})$ on the wave number~$\textbf{\textit{n}}$:
\begin{equation}\label{DR_exp}
{\omega(\textbf{\textit{n}})}\sim{\mu}\;k^2(\textit{\textbf{n}}) \qquad\dots\quad\mbox{dim.\ reduction}
\end{equation}
differs from our form:
\begin{equation}\label{GO_exp}
{\omega(\textbf{\textit{n}})}\sim\frac{k^2(\textit{\textbf{n}})}{\sqrt{k^2(\textit{\textbf{n}})+{m}^2}}  \qquad\dots\quad\mbox{GO proposal},
\end{equation}
where the wave momentum $k$ above fulfils
\begin{equation}\label{k}
k^2(\textit{\textbf{n}})=2\sum_i\left(1-\cos\frac{2\pi n_i}{L}\right).
\end{equation}
We therefore fitted our data for $\omega(\textbf{\textit{n}})$ at each coupling $\beta$ by the following two-parameter forms:
\begin{equation}\label{fits}
{\omega(\textbf{\textit{n}})}=\left\{
\begin{array}{l c l}
{a}+{b}k^2(\textit{\textbf{n}}) & \dots & \mbox{dim.\ reduction},\\[2mm]
\displaystyle\frac{{c}k^2(\textit{\textbf{n}})}{\sqrt{k^2(\textit{\textbf{n}})+{m}^2}} & \dots & \mbox{GO proposal}. \\
\end{array}\right.
\end{equation}
The result at $\beta=2.5$ ($30^4$ lattice) is displayed in figure \ref{omega_2_5_l30_c} (green dashed line for DR, blue dotted one for our proposal). Both forms fit the data
quite well at low plane-wave momenta, none of them is satisfactory for larger momenta. The situation at other gauge couplings is similar.

%
%\begin{center}
%\end{center}
%

Considerable improvement at all couplings is achieved by adding another parameter $d$ to our form:
\begin{equation}\label{best}
{\omega(\textbf{\textit{n}})}=
\frac{{c}k^2(\textit{\textbf{n}})}{\sqrt{k^2(\textit{\textbf{n}})+{m}^2}}\left[1+{d}k(\textit{\textbf{n}})\right],
\end{equation}
see the red solid line in figure \ref{omega_2_5_l30_c}. This would correspond, in the continuum limit, to adding a~term 
\begin{equation}\label{addition}
d\;\left(\frac{-{\cal{D}}^2-\lambda_0}{{(-{\cal{D}}^2-\lambda_0)+m^2}}\right)^{1/2}
\end{equation}
to the adjoint kernel $\mathfrak{K}$ that appears in the VWF (\ref{PsiGO3D}).

For constant configurations with small amplitude the DR and GO forms coincide. For consistency between our data for NAC and APW configurations, the value of $\mu_\mathrm{NAC}$ determined from sets of NAC configurations should agree with the appropriate combination $(2c_\mathrm{APW}/m_\mathrm{APW})$ of parameters obtained for abelian plane waves. Our results clearly pass this check quite success\-fully, both values converge to each other at large $\beta$, as exemplified in figure~\ref{NAC_vs_APW_c}.

\begin{figure}[t!]
\begin{minipage}[t]{18pc}
\fbox{\includegraphics[width=17.5pc]{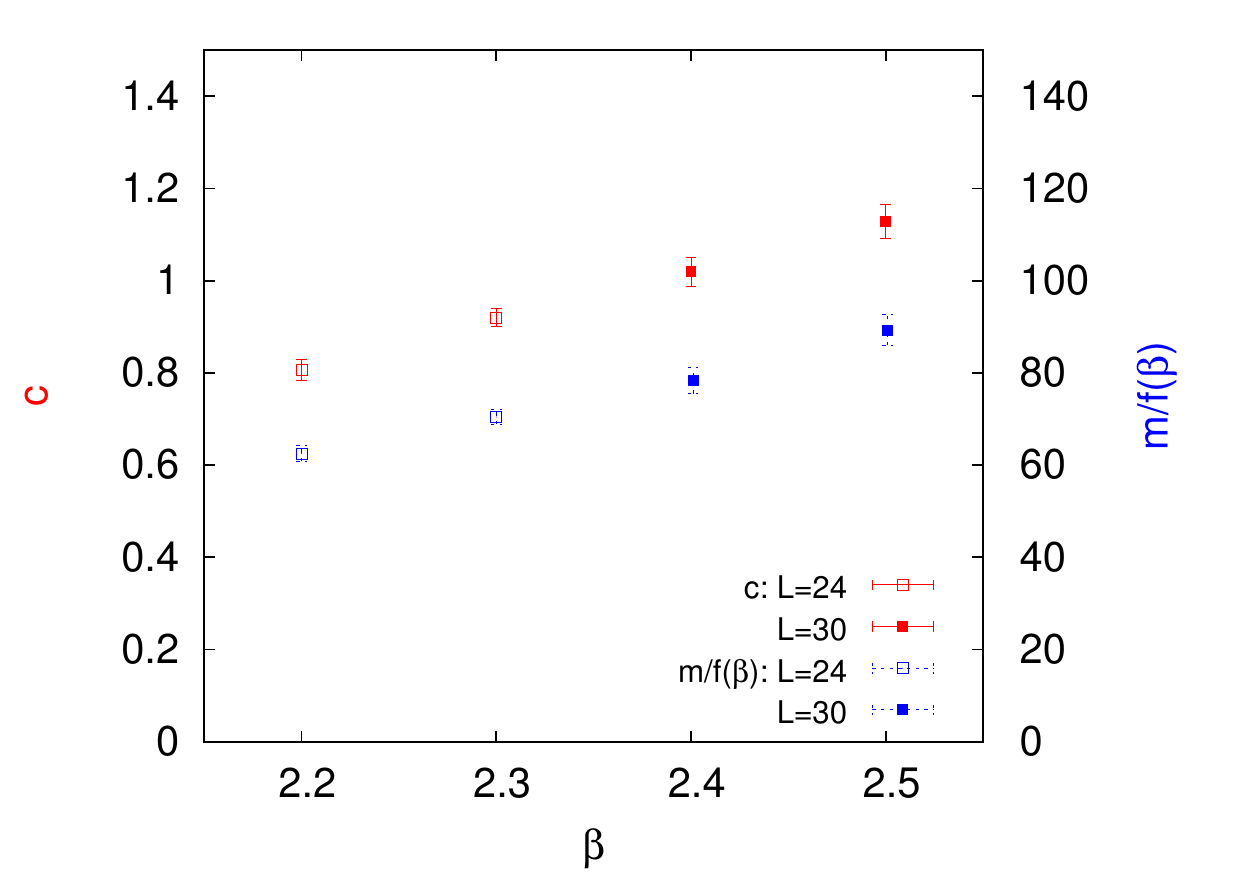}}
\caption{\label{cm_vs_beta_c}Parameter $c$ (red) and rescaled mass $m/f(\beta)$ (blue) of the best fit  (\ref{best}) vs.\ $\beta$.}
\end{minipage}
\hspace{1.5pc}%
\begin{minipage}[t]{18pc}
\fbox{\includegraphics[width=17.5pc]{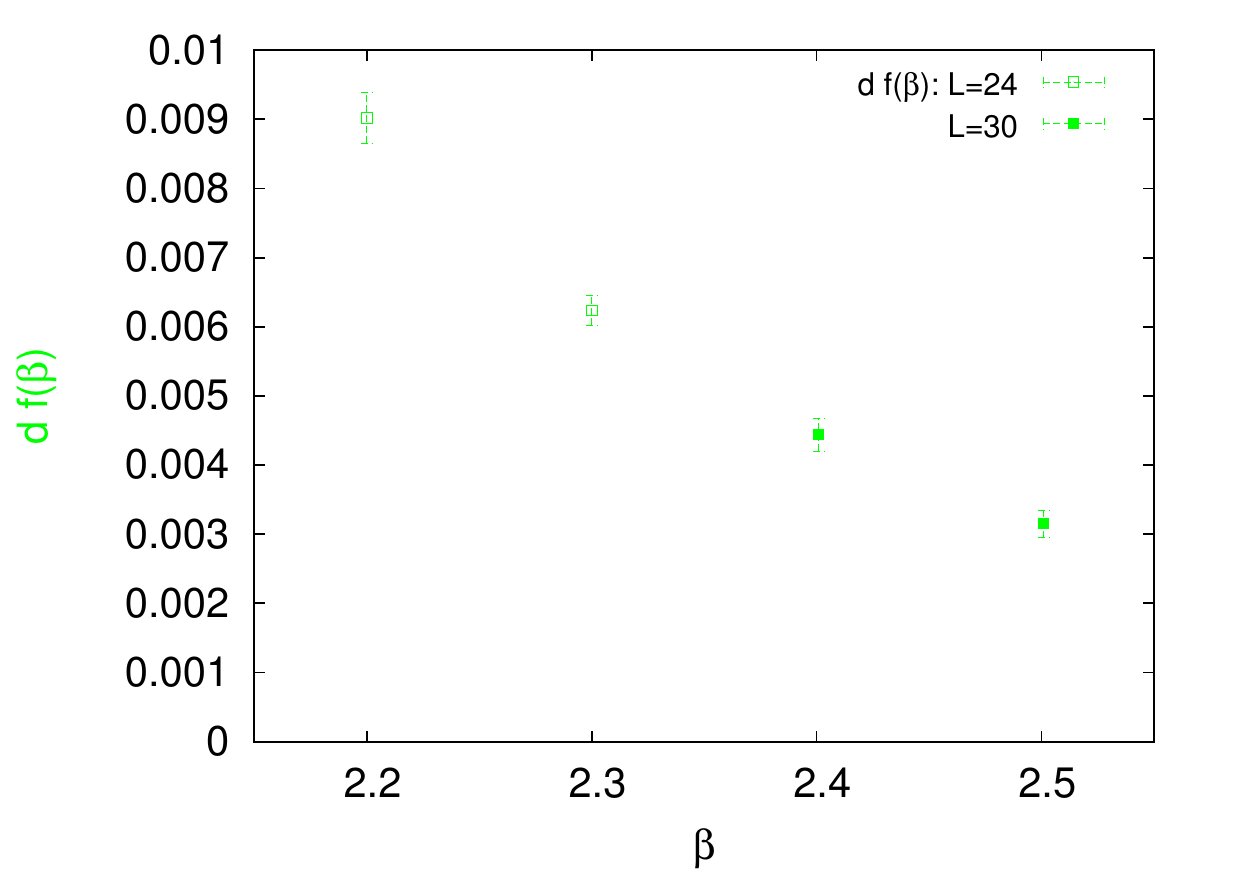}}
\caption{\label{d_vs_beta_c}Rescaled parameter $d f(\beta)$ (green points) of the best fit  (\ref{best}) vs.\ $\beta$.}
\end{minipage}
\end{figure}

Finally, the parameters of the best fit (\ref{best}), $c$, $m$ and $d$, if they correspond to some physical quantities in the continuum limit, should scale correctly when multiplied by the appropriate power of the asymptotic-freedom function $f(\beta)$ in (\ref{f_AF}). While the scaling of the ratio $(2c/m)$ multiplied by $f(\beta)$ has already been seen almost perfect in figure \ref{NAC_vs_APW_c}, it is not so convincing for individual parameters $c$ and $m/f(\beta)$, though their growth in the range of $\beta=2.2 \div 2.5$ is not large (see figure \ref{cm_vs_beta_c}) and one may optimistically hope that it will level off at still higher couplings. On the other hand, the parameter $d$ multiplied by $f(\beta)$ falls down quite rapidly in the same region (figure \ref{d_vs_beta_c}). The data thus suggest that the physical value of $d$ might vanish in the continuum limit. This signals a possibility that the form (\ref{PsiGO3D}) of the VWF might be recovered in the continuum limit.

%
%\begin{center}
%\end{center}
%

\section{Open end of the romance; roads to take}
Let me summarize what has been achieved until now, both positive (\includegraphics[height=0.5cm]{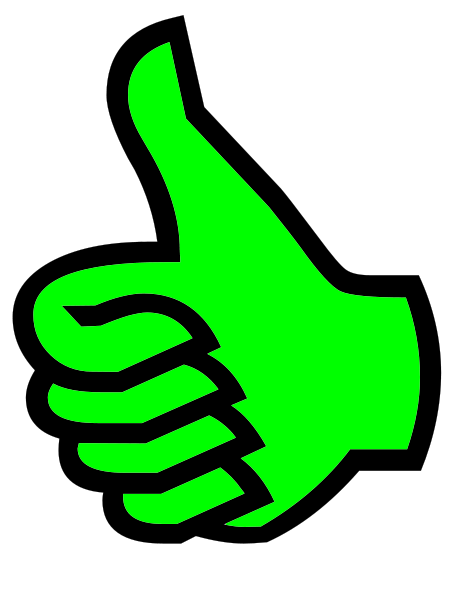}) and negative (\includegraphics[height=0.5cm]{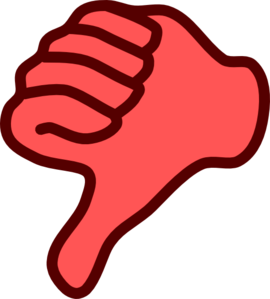}) points:

\bigskip 
\noindent\begin{minipage}[t]{18pc}
\begin{itemize}
\item[{\includegraphics[height=0.5cm]{up.png}}] We have proposed an approximate form of the SU(2) YM vacuum wave functional that looks good in $D=2+1$, somewhat worse in $(3+1)$ dimensions.
\end{itemize}
\end{minipage}\hspace{1.5pc}%
\begin{minipage}[t]{18pc}
\begin{itemize}
\item[{\includegraphics[height=0.5cm]{down.png}}] In $(3+1)$ dimensions, a method of generating configurations distributed according to the proposed vacuum wave functional is not (yet?) available.
\end{itemize}\end{minipage} 

\bigskip
\noindent\begin{minipage}[t]{18pc}
\begin{itemize}
\item[{\includegraphics[height=0.5cm]{up.png}}] The method of Greensite and Iwasaki allows one to compute numerically (on a lattice) relative probabilities of various gauge-field configurations in the YM vacuum.
\end{itemize}
\end{minipage}\hspace{1.5pc}%
\begin{minipage}[t]{18pc}
\begin{itemize}
\item[{\includegraphics[height=0.5cm]{down.png}}] This relative-weight method is only applicable to compute weights of configurations rather close in configuration space.
\end{itemize}\end{minipage} 

\bigskip
\noindent\begin{minipage}[t]{18pc}
\begin{itemize}
\item[{\includegraphics[height=0.5cm]{up.png}}] For non-abelian constant configurations and for long-wavelength abelian plane waves the measured probabilities are consistent with the dimensional-reduction form, and the coefficients $\mu$ for these sets agree.
\end{itemize}
\end{minipage}\hspace{1.5pc}%
\begin{minipage}[t]{18pc}
\begin{itemize}
\item[{\includegraphics[height=0.5cm]{down.png}}] Neither the dimensional-reduction form, nor our proposal for the vacuum wave functional  describe data satisfactorily for larger plane-wave momenta.
\end{itemize}\end{minipage} 

\bigskip
\noindent\begin{minipage}[t]{18pc}
\begin{itemize}
\item[{\includegraphics[height=0.5cm]{up.png}}] Numerical data for test configurations are nicely described by a natural modification of our proposal, and the correction term seems to vanish in the continuum limit.
\end{itemize}
\end{minipage}\hspace{1.5pc}%
\begin{minipage}[t]{18pc}
\begin{itemize}
\item[{\includegraphics[height=0.5cm]{down.png}}] No configurations tested so far, neither non-abelian constant nor abelian plane waves, can be considered typical. They are not in any sense true representatives of fields inhabiting the YM vacuum.
\end{itemize}\end{minipage} 

\bigskip
The investigation reported in this talk could be extended in various ways:

\medskip
-- One should compute, by the relative-weight method, weights of more realistic, ``typical'' configurations. One can \textit{e.g.\/}
generate ensembles of configurations of the full YM theory, take link matrices from a random time slice, make their Fourier de\-com\-posit\-ion, switch on/off or modify individual momentum modes, and compare the obtained momentum dependence of relative weights of configurations with that following from our, or some other, Ansatz for the VWF.

\medskip
-- It is most desirable to find a way of generating field configurations distributed according to (the square of) the VWF. What
was possible in $D=2+1$ dimensions, is made much more complicated by the Bianchi constraint in $(3+1)$.

\medskip
-- If one achieved progress in the above points and gained more evidence for the proposed form of the vacuum wave functional in $(3+1)$ dimensions, a number of other questions would call for an answer, \textit{e.g.\/}: What are the dominant field configurations in the YM vacuum? Where do colour screening and $N$-ality dependence arise from? \dots

\medskip
One can just hope, on the crooked road towards understanding the QCD vacuum wave functional, not to run into a ``Dead End'' sign.

\ack
I~am~most grateful to {Jeff Greensite} for collaboration on many topics, including those covered in the present talk, and to J\'an Pi{\v{s}}\'ut, who initiated my interest in this problem many years ago.
I would also like to thank organizers of \emph{DISCRETE 2014}, in particular {Joannis Papavassiliou}, the convener of its session on strongly coupled gauge theories, for inviting me to this interesting conference in solemn edifices of King's College London. I thank Ralf Hofmann for inviting me to the \textit{4th Winter Workshop on Non-Perturbative Quantum Field Theory} in Sophia Antipolis.
My research is supported by the {Slovak Research and Development Agency} under Contract No.\ APVV--0050--11, and by the {Slovak Grant Agency for Science}, Project VEGA No.~2/0072/13.

\appendix
\section{Bird's eye view of the YM theory on the lattice\footnote{For more details see \textit{e.g.\/} the textbook \cite{Gattringer:2010zz}.}}\label{A}

\begin{figure}[h!]
%\centerline
\begin{minipage}[t]{4.5pc}
~%\caption{\label{lattice}Building blocks of pure YM theory on a lattice.}
\end{minipage}
\fbox{{\includegraphics[height=6cm,trim=75 0 0 0,clip,angle=270]{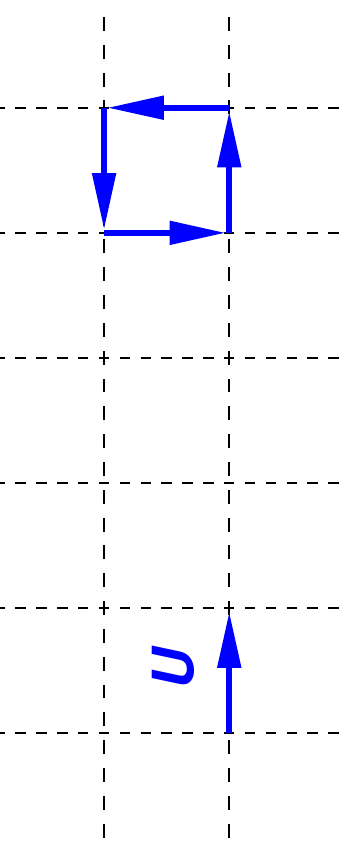}}}
\hspace{2pc}%
\begin{minipage}[t]{12pc}
\caption{\label{lattice}Building blocks of pure Yang--Mills theory on a lattice.}
\end{minipage}
\end{figure}

In the compact lattice formulation, gauge fields $\mathbf{A}_\mu(x)$ are represented by link matrices $U_\mu(x)$ (see figure \ref{lattice}):
\begin{equation}\label{U}
\mathbf{A}_\mu(x)=A_\mu^a(x)\mathbf{T}^a(x)\quad\longrightarrow\quad U_\mu(x)=\exp[iga\mathbf{A}_\mu(x)].
\end{equation}
  The field strengths $\mathbf{F}_{\mu\nu}(x)$ are related to products  $U_{\mu\nu}(x)$ of link matrices along a plaquette:
\begin{equation}\label{Umunu}
\mathbf{F}_{\mu\nu}(x)=F_{\mu\nu}^a(x)\mathbf{T}^a(x)\quad\longrightarrow\quad 
U_{\mu\nu}(x)= U_\mu(x) U_\nu(x+\hat{\mu})U^\dagger_\mu(x+\hat{\nu}) U^\dagger_\nu(x).
\end{equation}
%

%\begin{center}
%\end{center}
%

\noindent The simplest form that replaces the euclidean SU($N$) YM action
\begin{equation}\label{S_E}
{\cal S}=\frac{1}{2}\int d^4x\;\mbox{Tr}\;
\left[\mbox{\textbf{F}}_{\mu\nu}(x)\mbox{\textbf{F}}_{\mu\nu}(x)\right]
\end{equation}
is the Wilson action
\begin{equation}\label{S_W}
{\cal S}_W= a^4\;\beta
\sum_P\left[1-\frac{1}{N}\mbox{Re}\;\mbox{Tr}\;U_P\right]\quad\mbox{with}\quad
\beta=2N/g^2.
\end{equation}
Vacuum expectation values
\begin{equation}\label{VEV}
\langle 0 \vert {\widehat{{Q}}}\vert 0\rangle =
\displaystyle\int [dU]\;\mathbf{\Psi}^*_0[U]\;\widehat{{Q}}\;
\mathbf{\Psi}_0[U]
\end{equation}
are given by path integrals
\begin{equation}\label{VEV_PI}
\langle {\widehat{{Q}}}\rangle =
\displaystyle\frac{1}{Z}\int [dU_\mu(x)]\;{{Q}}[U]\;\exp(-{\cal{S}}_W[U]).
\end{equation}
In the numerical Monte Carlo simulation one computes in fact:
\begin{equation}\label{VEV_MC}
\langle {\widehat{{Q}}}\rangle \approx
\displaystyle\frac{1}{N_\mathrm{conf}}\sum_{i=1}^{N_\mathrm{conf}}Q[\{{\cal{C}}_i\}],
\end{equation}
an average over a (large) number $N_\mathrm{conf}$ of gauge-field configurations $\{{\cal{C}}_i\}$ distributed according to the probability distribution $\sim \exp(-{\cal{S}}_W[U])\sim\vert\mathbf{\Psi}_0[U]\vert^2$.

\section{Recursion method for simulation of the vacuum wave functional}\label{B}
Let us define a probability distribution for gauge fields $\mathbf{A}$ in a background of a second, independent configuration $\mathbf{A}'$:
\begin{equation}\label{probability}
{\cal P}[\mathbf{A};\mathfrak{K}[\mathbf{A}']]={\cal N}\left[\mathbf{A}'\right]\exp\left[-\displaystyle\int d^2x\;d^2y\; 
B^a(x;\mathbf{A})\;{\mathfrak{K}_{xy}^{ab}[\mathbf{A}']}\;B^b(y;\mathbf{A})\right].
\end{equation}
where $\mathbf{A}$ and $\mathbf{A}'$ are fixed to a variant of axial gauge (Appendix B of~\cite{Greensite:2007ij}). If we assume that the variance of $\mathfrak{K}$ is small among thermalized configurations, we can approximate:
\begin{equation}\label{hypothesis}
{P[\mathbf{A}]}\equiv{\cal P}[\mathbf{A};\mathfrak{K}[\mathbf{A}]]\approx{\cal P}[\mathbf{A};\langle \mathfrak{K}\rangle]
\approx\int d\mathbf{A}'\;{\cal P}[\mathbf{A};\mathfrak{K}[\mathbf{A}']]\; {P[\mathbf{A}']}.
\end{equation}
Then the probability distribution can be generated by solving (\ref{hypothesis}) iteratively:
\begin{eqnarray}
P^{(1)}[\mathbf{A}]&=&{\cal P}\left[\mathbf{A};\mathfrak{K}[0]\right],\\
P^{(k+1)}[\mathbf{A}]&=&\int d\mathbf{A}'\;{\cal P}\left[\mathbf{A};\mathfrak{K}[\mathbf{A}']\right]\; P^{(k)}[\mathbf{A}'].
\end{eqnarray}
A block diagram of a practical implementation of this recursion procedure is shown below:\\

%
%\centerline{{\fbox{\includegraphics[trim={-10 -10 -10 -10},width=18pc]{recursion.eps}}}}
%\begin{center}
\centerline{{\fbox{\includegraphics[trim={-10 -10 -10 -10}]{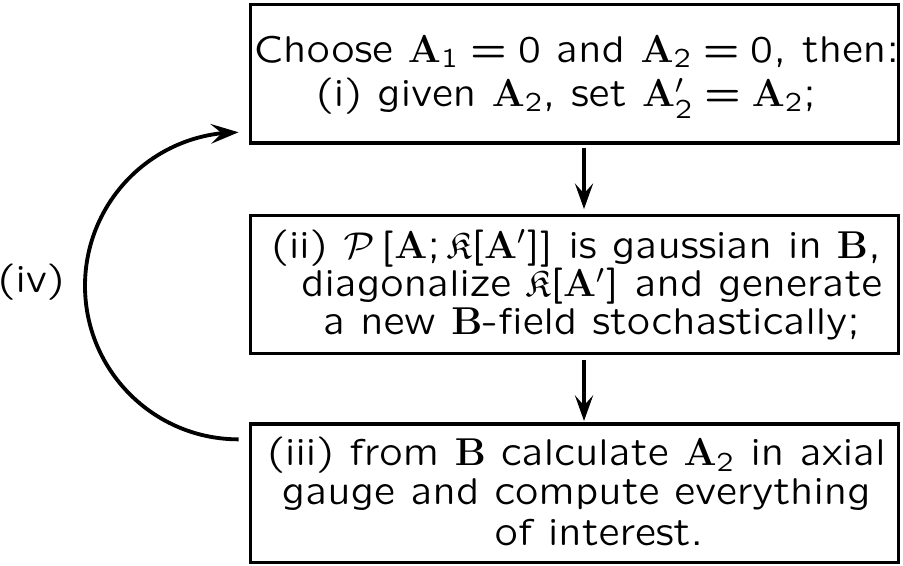}}}}
%\end{center}

\bigskip\noindent
In the case of the kernel appearing in the vacuum wave functional (\ref{PsiGO}), the procedure converges quite rapidly, typically in ${\cal{O}}(10)$ above cycles. The hypothesis about small changes of $\mathfrak{K}$ among equilibrated configurations is confirmed \textit{a posteriori\/} by the absence of large fluctuations of the spectrum of $\mathfrak{K}$ for individual recursion lattices.
 
%\section*{References}
\raggedright
%\bibliography{DISCRETE2014}
\providecommand{\newblock}{}

\end{document}